\documentclass[journal]{IEEEtran}

\usepackage{xcolor}
\usepackage{cite}
\ifCLASSINFOpdf
  \usepackage[pdftex]{graphicx}
\else
\fi
\begin{document}
\title{I-nteract 2.0: A Cyber-Physical System to Design 3D Models using Mixed Reality Technologies and Deep Learning for Additive Manufacturing}

\author{Ammar~Malik,
        Hugo~Lhachemi,
        and~Robert~Shorten% <-this % stops a space
\thanks{Ammar Malik is with the Department
of Electrical and Electronic Engineering, University  College  Dublin,  Dublin,  Ireland.  e-mail:  ammar.malik@ucdconnect.ie. Hugo Lhachemi is with L2S, CentraleSup{\'e}lec, 91192 Gif-sur-Yvette, France.  e-mail:  hugo.lhachemi@centralesupelec.fr. Robert Shorten is with the Dyson School of Design Engineering, Imperial College London, London, United Kingdom. e-mail: r.shorten@imperial.ac.uk.}% <-this % stops a space
\thanks{This publication has emanated from research supported in part by a research grant from Science Foundation Ireland (SFI) under grant number 16/RC/3872 and is co-funded under the European Regional Development Fund and by I-Form industry partners.}% <-this % stops a space
}

\maketitle

\begin{abstract}
I-nteract is a cyber-physical system that enables real-time interaction with both virtual and real artifacts to design 3D models for additive manufacturing by leveraging on mixed reality technologies. This paper presents novel advances in the development of the interaction platform I-nteract to generate 3D models using both constructive solid geometry and artificial intelligence. The system also enables the user to adjust the dimensions of the 3D models with respect to their physical workspace. The effectiveness of the system is demonstrated by generating 3D models of furniture (e.g., chairs and tables) and fitting them into the physical space in a mixed reality environment.
\end{abstract}


% \fi
\IEEEpeerreviewmaketitle

\section{Introduction}

\begin{figure}
    \centering
    \includegraphics[width=3.5in]{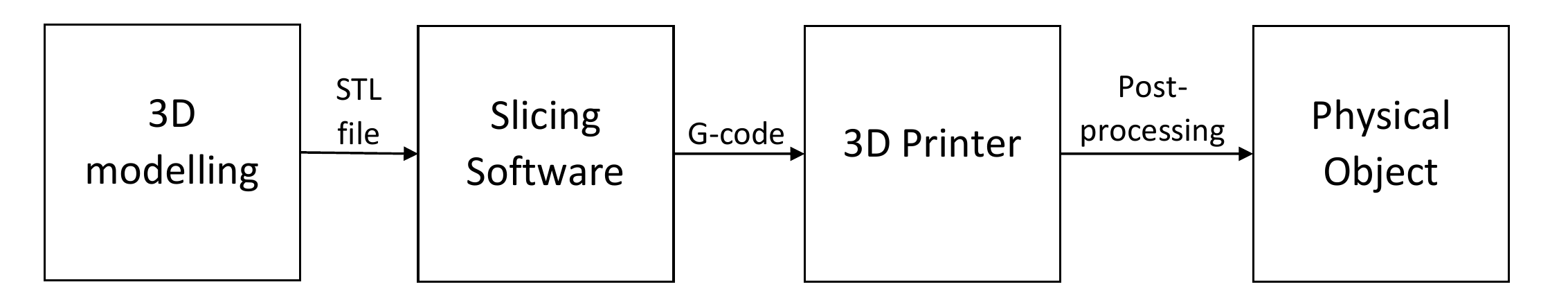}
    \caption{AM workflow.}
    \label{workflow}
\end{figure}

%T}{he} rapid pace of technological developments in the areas of cyber-physical systems (CPS), internet of things (IoT), additive manufacturing (AM), mixed reality (MR), cloud computing, robotics, and machine learning (ML) is leading us towards the vision of the fourth industrial revolution\cite{zhong2017intelligent}. A revolution that is driving the global trend towards edge manufacturing and the prosumer driven society. A society in which a customer is not only consuming but is also involved in the design process of manufacturing a product tailored according his/her individual demands. 
\IEEEPARstart{} Additive manufacturing (AM) has emerged during the last decade as a key enabling technology poised to deeply transform manufacturing\cite{shorten2020industry, RN118, RN6}. AM, also known as 3D printing, rapid prototyping, or generative manufacturing, refers to the process of depositing successive thin layers of materials upon each other in precise geometric shapes based on 3D model files to manufacture three-dimensional physical objects\cite{attaran2017rise}. A general workflow of AM is depicted in Fig \ref{workflow}. It starts with the three dimensional virtual model of the desired product designed via a computer-aided design (CAD) tool or obtained from 3D scanning. The 3D model (STL file) is sliced into layers and converted into specific instructions (g-code) for the 3D printer using a slicing software. Then the 3D printer executes the instructions to build the physical object layer upon layer. Finally, the post-processing is done either to remove support structures or to give the finishing touch to the 3D printed product. This workflow is sub-optimal due to a lack of feedback between the AM process and the 3D modelling. Indeed, in such a workflow, testing of the designed 3D model for the desired functionality is postponed to the end of the printing process. Hence, the entire loop is reiterated through a trial-error procedure until the desired results are achieved, making the design process costly and time-consuming. Moreover, most CAD design software not only require professional training but also restrain the design of 3D virtual models to 2D interfaces, making the design process unintuitive and cumbersome for the non-technical consumers\cite{RN87,RN53}. In this context, innovations in the design of CPS and technological advancements in its supporting tools (IoT, MR, robotics, ML) are playing an important role for the widespread adoption of AM by the general public as well as the industry\cite{lhachemi2019augmented}. 

I-nteract\cite{malik2020nteract} is a CPS that enables the user to interact with both the virtual as well as the physical objects (deformable and non-deformable) simultaneously in a visio-haptic mixed reality (VHMR) environment. The system streamlines the AM process by allowing the user to generate digital twins of the real objects and to test the properties of the designed virtual model in response to human and physical objects stimuli prior to printing. Such innovations in the development of CPS are not only enabling the development of intuitive interfaces for human-machine interactions (human-in-the-loop) \cite{RN1,RN13,RN45} but they also provide innovative monitoring solutions to improve the build quality of the product\cite{malik2019, RN77}.

In comparison to traditional (subtractive and formative) manufacturing, AM allows the manufacturing of complex geometries without using traditional dies, molds, milling, and machining which are expensive and time consuming for mass customization\cite{klahn2015design}. This advantage over traditional manufacturing makes AM a key enabler in producing moderate to mass quantities of products that can be customized individually for personal fabrication\cite{attaran2017rise}. Although existing solutions \cite{malik2020nteract,RN1,RN13,RN45} provide innovative interfaces to bridge the gap between the consumer, the designer, and the production using AM but only allow either modifications in the existing 3D model or 3D scanning of an existing real object. In this context, there is a need for interfaces that, along with providing an immersive experience in the three-dimensional workspace, also enable the non-technical users to design 3D models from scratch with minimum effort. Constructive solid geometry (CSG) and ML can play a significant role to achieve this objective. CSG, also known as building block geometry, offers simple, precise, and concise methods for generating 3D models\cite{requicha1977constructive}. Recent developments in the generative networks \cite{liu2019soft, kato2018neural, yan2016perspective, sinha2017surfnet, lin2017learning, dosovitskiy2016learning, wang2017shape, wu2016learning, fan2017point, gadelha20173d}, a subbranch of deep learning (DL), provide an effective solution to automate the parts of the design process that require expert knowledge for generating 3D models. 

In this paper, we present I-nteract 2.0, an advanced development of its predecessor I-nteract, that enables generative CAD in MR by taking advantage of CSG and DL. I-nteract 2.0 also exploits the immersive feature of MR by enabling the user to adjust the dimensions of the virtual model with respect to the design constraints in the physical workspace.

The remainder of this paper is structured as follows. Related works are presented in Section II. After a general description of the system in Section III, the methods to generate 3D models using I-nteract 2.0 are illustrated in Section IV. Results are reported in Section V. Finally, concluding remarks are provided in Section VI.

% process of additive manufacturing (generation, modification and, monitoring), capabilities of I-nteract 1.0 in modification and monitoring

\section{Related Works}
The technological advancements in the areas of MR, robotics, computer vision, and ML has already enabled the development of many intuitive and realistic interfaces for humans to interact with both the physical and digital world in real time. Further, in recent years, extensive research has been devoted to improve the real-time representation of the virtual world in users' physical environment using these innovative technologies \cite{lhachemi2019augmented}. The present section focuses on the research endeavors of such novel interfaces in the context of 3D modelling for AM.  

Window-Shaping \cite{RN91} is an AR interface with the objective of integration of physical objects into the design process. The interface consists of a hand-held device to enable the user to perform sketch-based 3D modelling in reference to physical artifacts. Although window-shaping merges the digital and the physical worlds but provides 2D view of a three-dimensional workspace. Modern MR solutions remedy this either by stereoscopic projections or head-mounted displays (HMDs) which also allow the user to use the hands in three-dimensional space for interaction hence enabling a more immersive experience. MirageTable \cite{RN9}, a freehand interactive system utilises a depth camera, a curved screen, and a stereoscopic projector to provide a MR interface for 3D modelling using gestures. Interactive situated AR systems like HoloDesk\cite{RN13}, Holo TableTop\cite{RN45}, and MixFab\cite{RN1} provide intuitive interfaces to enable personal fabrication for non-technical designers. MixFab along with a depth camera for hand gestures detection and a MR display consists of a motorized turntable to enable 3D scanning of a physical object. The user then can use the scanned virtual model as a size or shape reference to design 3D models. Tangible interaction with intangible objects in an immersive augmented environment makes the experience more realistic which cannot be achieved by relying solely on visual feedback and gestures. For improving interactivity, interfaces like Surface Drawing\cite{RN22}, Twister\cite{RN18}, Digits\cite{RN16}, and NormalTouch and TextureTouch\cite{RN72} make use of additional hardware (such as haptic gloves) for force feedback to enable physical interaction with virtual artifacts. I-nteract\cite{malik2020nteract} is a VHMR system that comprises MR glasses for visual feedback, haptic glove for force feedback, and force sensors to enable dynamic interaction between human, physical, and virtual objects to streamline the design process for AM. 

ML, a subset of AI, is a powerful tool that enables the system to learn automatically from data without being explicitly programmed to perform a task. Researchers are actively involved in exploring innovative ways to integrate ML within the AM process. In recent years, ML has proven to be useful in improving product quality, optimizing manufacturing processes, and reducing costs \cite{meng2020machine}. DL, a subset of ML, has emerged as an active research area to enable generative design. Generative design is an iterative design exploration process that involves the automatic generation of design options to meet certain constraints. These options are presented to the designer to fine-tune. This automated generation makes it feasible for non-technical and inexperienced users to implement their ideas. Generative design has also been integrated into many commercially available CAD packages such as Ansys\footnote{https://www.ansys.com/}, Autodesk\footnote{https://www.autodesk.com/}, etc. Generative design framework generates outputs that are not only aesthetic but also satisfy engineering constraints.  Generative modelling is an active research area of DL that has a great potential to improve generative design \cite{oh2019deep}. Generative models although not yet used to its full potential to produce engineering designs\cite{oh2018design} but have already proven themselves to be immensely capable of inferring 3D shapes from 2D images. Variational autoencoders (VAEs)\cite{kingma2013auto} and generative adversial networks (GANs)\cite{goodfellow2014generative} are the two significant types of generative deep convolution neural networks (CNNs) that have been extensively researched to perform generative tasks\cite{liu2019soft, kato2018neural, yan2016perspective, sinha2017surfnet, lin2017learning, dosovitskiy2016learning, wang2017shape, wu2016learning, fan2017point, gadelha20173d}. 

With the technological advancements in MR technologies along with the democratization of 3D printers, generative modelling using deep neural networks (DNN) has emerged as a promising tool to generate 3D models for AM\cite{sinha2017surfnet}. 3D models have various form of representations which lead to different DNN architectures. Volumetric (voxelized), mesh, and point cloud are the most popular and widely used 3D model representations. Each representation has its own merits when used in generative modelling. Although volumetric representation enables the 3D CNNs, a direct extension of 2D CNNs, for 3D content generation but is computationally wasteful as most information of a 3D shape lies on the surface hence making the extra third dimension redundant. Mesh and point cloud representations provide compact encoding of shape information but suffer from dimensional variability per 3D shape sample that complicates the application of learning methods to infer 3D shapes from 2D images \cite{lin2017learning}. Generative modelling using template mesh deformation \cite{liu2019soft,kato2018neural,wang2018pixel2mesh} is an innovative solution to deal with this problem. As mesh representation (using triangular meshes) is predominantly used for 3D models representation both in MR and AM therefore the generative DNNs based on mesh representation of 3D models are more compatible to be integrated within the MR based AM design process. The common mesh representation based 3D file formats are OBJ and STL. These file formats contain information about the vertices and the faces of the triangles to estimate the 3D shapes. In the template mesh deformation method, the DNN learns the displacement in the position of the vertices to synthesize a 3D model with respect to the input image. In this method, the number of vertices and the faces remain constant which solves the inherent problem of dimensional variability per 3D shape sample in using mesh representation. The generation of a 3D model based on a single 2D image is termed as single-view mesh reconstruction in literature. To take advantage of the generative capability of the DNNs we have integrated SoftRas\cite{liu2019soft} based generative model with I-nteract\cite{malik2020nteract} for single-view mesh reconstruction in a MR environment.

% Generative DL is a vastly growing area of research for generating 3D models based on 2D images.

% Deep learning possess an immense potential in enabling generative design for AM.

% Among several challenges, 3D shape representation is the fundamental one that needs to be addressed for learning a generative model of 3D shapes from 2D images.

% MR for CAD model visualization and interactive validation only,
% AR and machine learning for the monitoring of the manufacturing process only

\section{System Overview}
\begin{figure}
    \centering
    \includegraphics[width=3.25in]{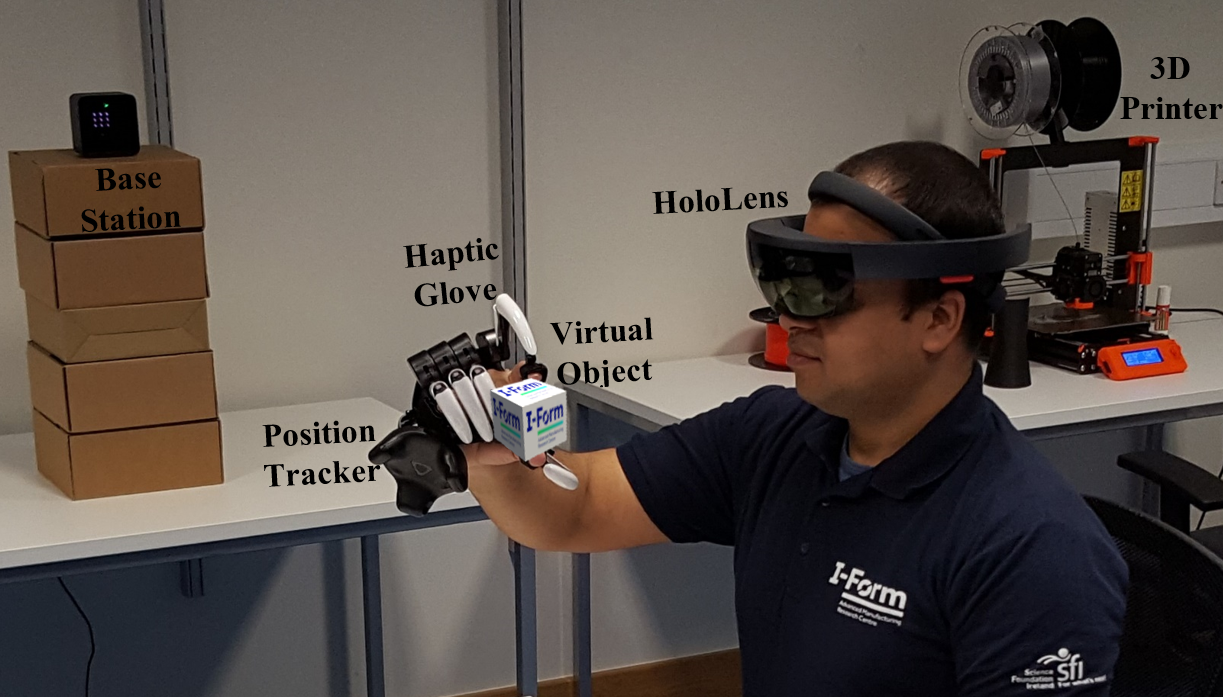}
    \caption{I-nteract \cite{malik2020nteract}.}
    \label{fig:sys}
\end{figure}
I-nteract utilises MR and haptic feedback to provide the user with an integrated visio-haptic experience to design 3D models for AM \cite{malik2020nteract}. I-nteract allows the designers to inspect and perfect virtual objects in real-time based on the interaction with other objects or humans prior to printing, and in this way streamlines the AM process. The system is built using MR smartglasses (HoloLens) for visual feedback, haptic gloves (Dexmo) for force feedback, and VIVE\footnote{https://www.vive.com/ca/vive-tracker/} hardware for global position tracking of the hand (glove). The objective is that I-nteract provides an intuitive novel MR interface to 3D scan a physical object and to measure its physical properties (such as elasticity) to generate a digital twin. The interaction of a user with a virtual object using I-nteract is illustrated in Fig.~\ref{fig:sys}.  

\begin{figure*}
    \centering
    \includegraphics[width=6in]{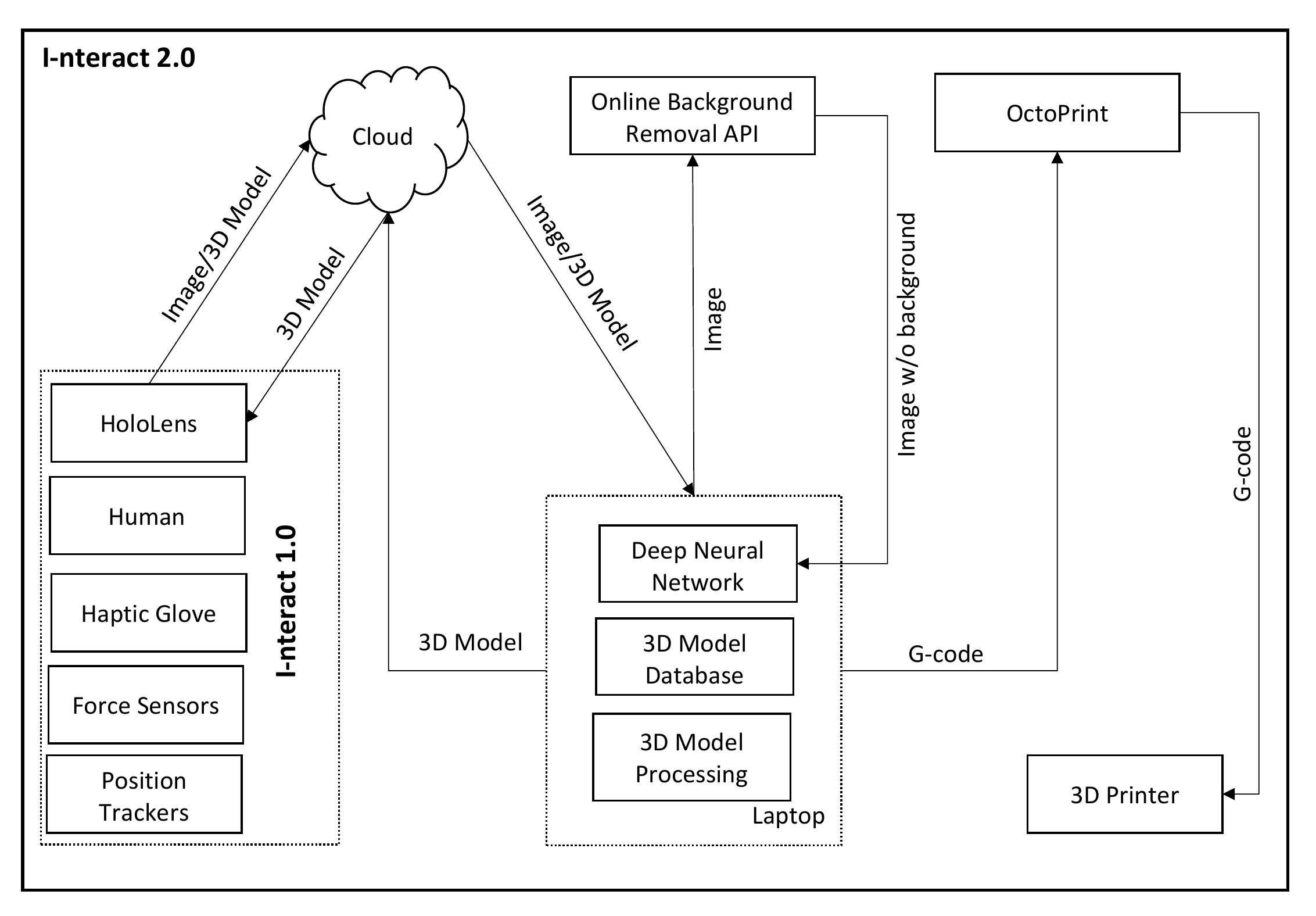}
    \caption{System architecture.}
    \label{fig:sys_arch}
\end{figure*}

\begin{figure}
    \centering
    \includegraphics[width=3.25in]{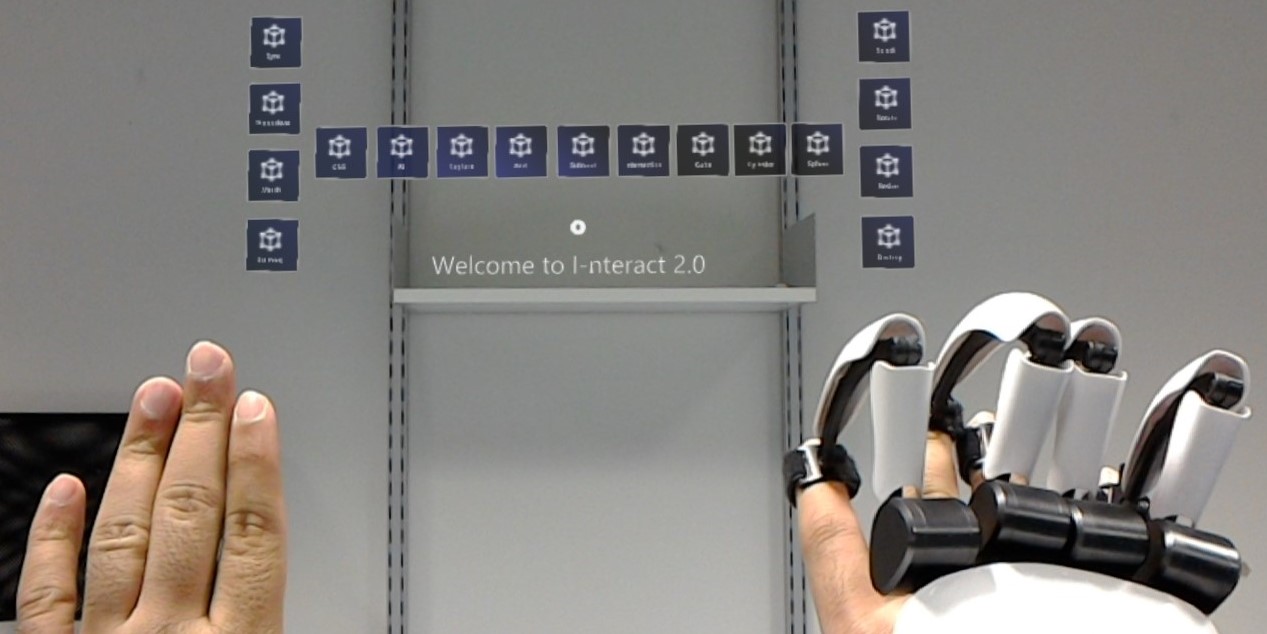}
    \caption{Interface for interaction.}
    \label{fig:Interface}
\end{figure}

In this paper, we present further developmental advances in the VHMR system (I-nteract) reported in \cite{malik2020nteract} by using CSG and AI to enable CAD in MR for AM. To the best of our knowledge, I-nteract~2.0 is the first VHMR system that enables generative AI based CAD in MR for AM. Integration with CSG allows the user to design 3D models from scratch using primitive 3D objects (such as cuboids, cylinders, spheres, etc.) and his/her creative skills in a MR environment. The CSG for creating meshes in MR using boolean operations has been adapted from \cite{CSG_Unity}. The AI network integration enables the user to generate 3D models automatically by taking pictures of the objects using HoloLens. The detailed system architecture that defines the flow of information between the different modules of I-nteract~1.0 can be found in Fig.~3 of \cite{malik2020nteract}. The updated system architecture of I-nteract~2.0 after integration with CSG and DNN is depicted in Fig.~\ref{fig:sys_arch}. As illustrated in Fig.~\ref{fig:sys_arch}, the image or the 3D model is sent to the cloud to be accessed by the HoloLens and the laptop. The 3D print controller application OctoPrint\footnote{https://octoprint.org/} has been used to send the 3D model to the printer PRUSA i3 MK3\footnote{https://www.prusa3d.com/}. The MR interface is shown in Fig.~\ref{fig:Interface}. The interface consists of a hand with glove, a hand without glove, virtual buttons, and voice commands. The hand with glove can be used to translate, rotate, and resize the 3D model while getting force feedback. The hand without glove can be used to utilize the built-in interface of the HoloLens such as moving the 3D model and pressing the virtual buttons. The user can control the interface using voice commands, gestures (from the hand without haptic glove), and hand motions as well as finger motions (of the hand with haptic glove). The respective functions and the associated voice commands of the virtual buttons are detailed in Tab.~\ref{tab:buttons}.

\begin{table*}[]
\centering
\caption{Virtual buttons, voice commands, and their respective functions. }
\label{tab:buttons}
\begin{tabular}{|c|l|}
\hline
\textbf{Virtual Button / Voice Command} & \multicolumn{1}{c|}{\textbf{Function}}                                      \\ \hline
Sync                                    & To sync the position of the real hand with glove and the virtual hand model.           \\ \hline
Cube                                    & To start drawing a cuboid.                                                  \\ \hline
Sphere                                  & To start drawing an ellipsoid.                                              \\ \hline
Cylinder                                & To start drawing a cylinder.                                                \\ \hline
Add                                     & Union of the primitive shapes.                                              \\ \hline
Subtract                                & Difference of the primitive shapes.                                         \\ \hline
Intersect                               & Intersection of the primitive shapes.                                       \\ \hline
Select                                  & To select the 3D model.                                                     \\ \hline
Capture                                 & To take an image using HoloLens.                                            \\ \hline
Dimension                               & To display the dimensions of the selected 3D model.                          \\ \hline
Resize                                  & To resize the selected 3D model.                                            \\ \hline
Match                                   & To find the best possible match of the selected 3D model from the database. \\ \hline
Print                                   & To send the 3D model to the printer.                                        \\ \hline
\end{tabular}
\end{table*}

\section{Generating 3D models in a mixed reality environment}
The conventional graphical user interface (GUI) for 3D modelling renders the virtual 3D world on a 2D computer screen. This makes the use of mouse and keyboard to locate and place virtual objects in a 3D environment unintuitive and difficult for inexperienced users. Also, most contemporary CAD-based software demands strong technical background which makes it even more difficult for non-technical consumers to participate in the design process \cite{lhachemi2019augmented}. In this context, there is a clear need for developing innovative interfaces that not only take advantage of MR technologies for interacting with 3D models in a three-dimensional environment but also enable generative CAD in MR and utilise ML to automate the parts of the design process that require expert knowledge. I-nteract is a CPS that provides a framework to develop such intuitive and automated interfaces for assembling, creating, interacting, modifying, positioning, and shaping 3D models within a three-dimensional environment. Built upon I-nteract, I-nteract~2.0 uses the generative functionalities of CSG and DL to enable the user to create 3D models from 3D primitive shapes as well as to automate the generation of the 3D models based on 2D images. Taking advantage of the immersive feature of MR, I-nteract 2.0 also allows the user to modify dimensions of a 3D model with respect to the physical workspace. 

\subsection{Generating 3D models using constructive solid geometry}
\label{CSG}
\begin{figure*}
\centering
\includegraphics[width=6.5in]{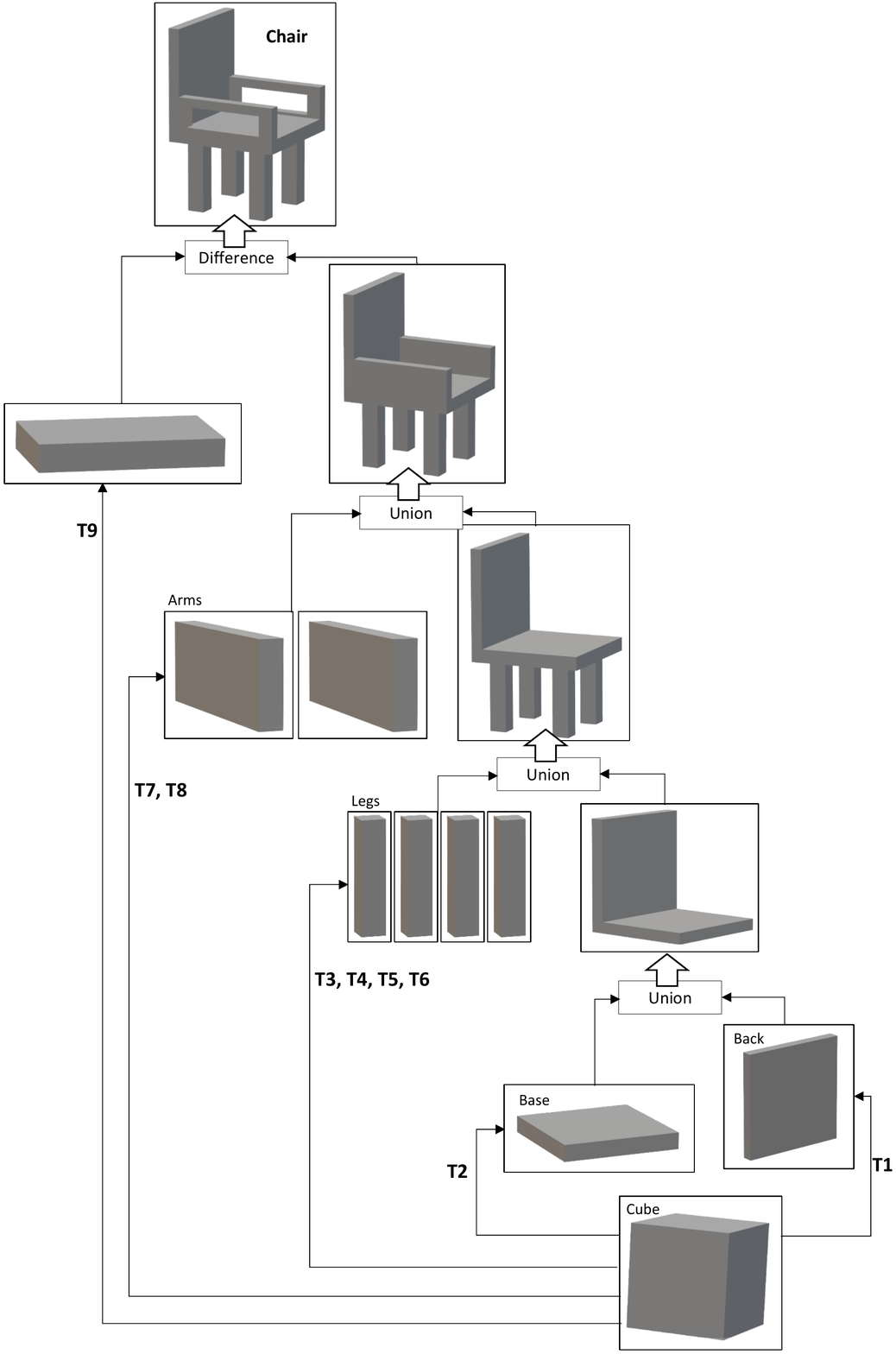}
\caption{Generating 3D model of a chair using CSG.}
\label{CSG_fig}
\end{figure*}

\begin{figure*}
\centering
\includegraphics[width=7in]{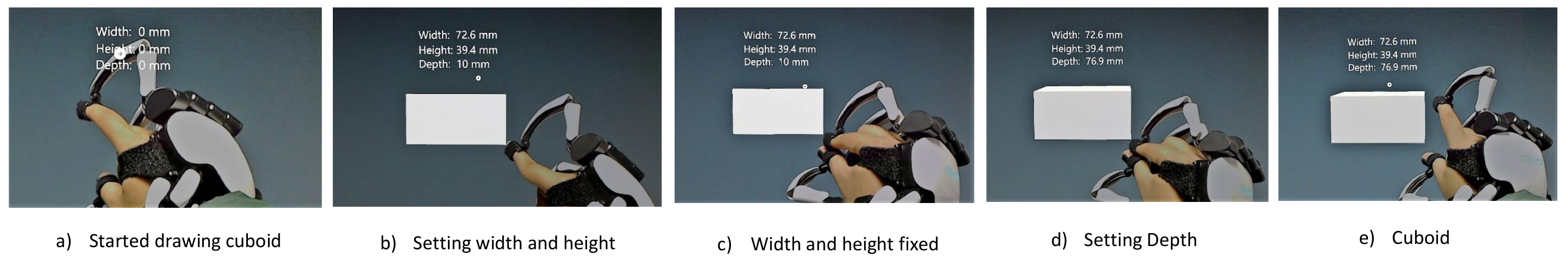}
\caption{Drawing a cuboid in MR using I-nteract 2.0.}
\label{CSG_Step2}
\end{figure*}

Constructive solid geometry (CSG), used in solid modelling, allows the user to construct complex 3D models by using boolean set operations (e.g., union, difference, and intersection) on simple building blocks (e.g., cubes, cylinders, and spheres) called primitives. We have utilised CSG in the system to enable the user to intuitively design 3D models in a MR environment from primitive shapes. An example of creating a chair using CSG is illustrated in Fig.~\ref{CSG_fig}. Tab.~\ref{CSG_tab} depicts the transformations applied to the cube in the example, shown in Fig.~\ref{CSG_fig}, to translate, rotate, and resize the primitive shapes. The position, rotation, and scale vectors given in Tab.~\ref{CSG_tab} are the same vectors that are used in Unity\footnote{https://unity.com/} to transform a 3D model. The hand with the glove can be used to grab (in order to translate or rotate the model) or resize the virtual object in the 3D physical workspace. The hand without the glove can be used to translate the virtual object. This feature is useful when the user is using the other hand (with glove) to resize the virtual object so that the user can place and resize/rotate the virtual object simultaneously in the physical workspace by using both hands.
The procedure implemented to draw a 3D primitive shape using the hand with glove is illustrated by Fig.~\ref{CSG_Step2} (a-e). The procedure implemented to generate a 3D model using CSG is described below. 

\begin{figure}
\centering
\includegraphics[width=3.25in]{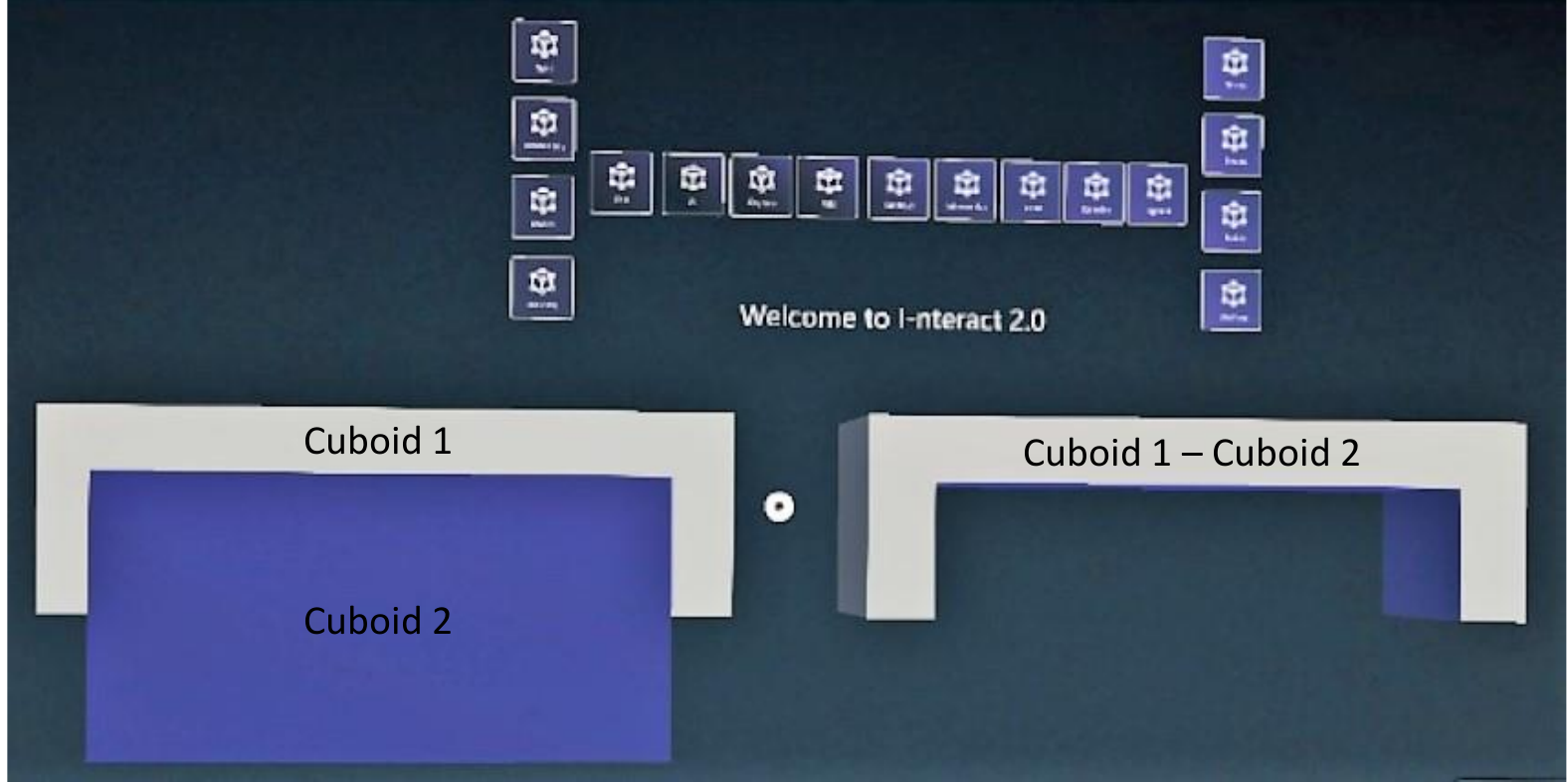}
\caption{Subtracting two cuboids in MR using I-nteract 2.0.}
\label{CSG_Sub}
\end{figure}

\begin{enumerate}
    \item Position the hand with glove in the physical workspace where the primitive shape is desired to be drawn. The index finger of the hand should be open as shown in Fig.~\ref{CSG_Step2} (a). Press the virtual button of the desire primitive shape (Cube, Sphere, Cylinder) or use the associated voice command, as described in Tab.~\ref{tab:buttons}, to start drawing the primitive shape. 
    \item After selecting the desired primitive shape, the width and the height of the primitive shape can be adjusted by moving the hand with glove in left/right (x) and up/down (y) direction respectively as shown in Fig.~\ref{CSG_Step2} (b). Close the index finger of the hand when done as shown in Fig.~\ref{CSG_Step2} (c).
    \item Move the hand in the forward/backward (z) direction to adjust the depth of the primitive shape as shown in Fig.~\ref{CSG_Step2} (d). Open the index finger of the hand with glove when done as shown in Fig.~\ref{CSG_Step2} (e).
    \item Repeat steps 1 to 3 to draw another primitive shape.
    \item Apply the transformations (translation or rotation) using hand with glove or (translation and rotation) using both hands to place primitive shapes at desired locations and orientations.
    \item Press the "Select" virtual button or use the associated voice command to select the primitive shapes.
    \item Press the virtual button of the desired boolean operation (Add, Subtract, Intersection) or use the associated voice command to apply the boolean operation (Union, Difference, Intersection respectively). Fig.~\ref{CSG_Sub} shows the subtraction of two cuboids in MR using I-nteract 2.0.
    \item Repeat steps 1 to 7 to generate a 3D model from the primitive shapes.
    
\end{enumerate}

\begin{table*}[]
\centering
\caption{Transformations to cuboid for CSG shown in Fig. \ref{CSG_fig}.}
\label{CSG_tab}
\begin{tabular}{|c|c|c|c|}
\hline
\textbf{Transformations} & \textbf{Position Vectors}                      & \textbf{Rotation Vectors}        & \textbf{Scale Vectors}                      \\ \hline
T1                       & \textless{}0, 0.044, 0.4\textgreater{}         & \textless{}0, 0, 0\textgreater{} & \textless{}0.1, 0.1, 0.01\textgreater{}     \\ \hline
T2                       & \textless{}0, 0, 0.355\textgreater{}           & \textless{}0, 0, 0\textgreater{} & \textless{}0.1, 0.012, 0.1\textgreater{}    \\ \hline
T3                       & \textless{}-0.03, -0.034, 0.386\textgreater{}  & \textless{}0, 0, 0\textgreater{} & \textless{}0.015, 0.07, 0.015\textgreater{} \\ \hline
T4                       & \textless{}-0.03, -0.034, 0.326\textgreater{}  & \textless{}0, 0, 0\textgreater{} & \textless{}0.015, 0.07, 0.015\textgreater{} \\ \hline
T5                       & \textless{}0.03, -0.034, 0.326\textgreater{}   & \textless{}0, 0, 0\textgreater{} & \textless{}0.015, 0.07, 0.015\textgreater{} \\ \hline
T6                       & \textless{}0.03, -0.034, 0.386\textgreater{}   & \textless{}0, 0, 0\textgreater{} & \textless{}0.015, 0.07, 0.015\textgreater{} \\ \hline
T7                       & \textless{}-0.045, 0.017, 0.3525\textgreater{} & \textless{}0, 0, 0\textgreater{} & \textless{}0.01, 0.035, 0.095\textgreater{} \\ \hline
T8                       & \textless{}0.045, 0.017, 0.3525\textgreater{}  & \textless{}0, 0, 0\textgreater{} & \textless{}0.01, 0.035, 0.095\textgreater{} \\ \hline
T9                       & \textless{}0, 0.0175, 0.3525\textgreater{}     & \textless{}0, 0, 0\textgreater{} & \textless{}0.12, 0.018, 0.07\textgreater{}  \\ \hline
\end{tabular}
\end{table*}

\subsection{Generating 3D models using deep learning}
\begin{figure}
\centering
\includegraphics[width=3.25in]{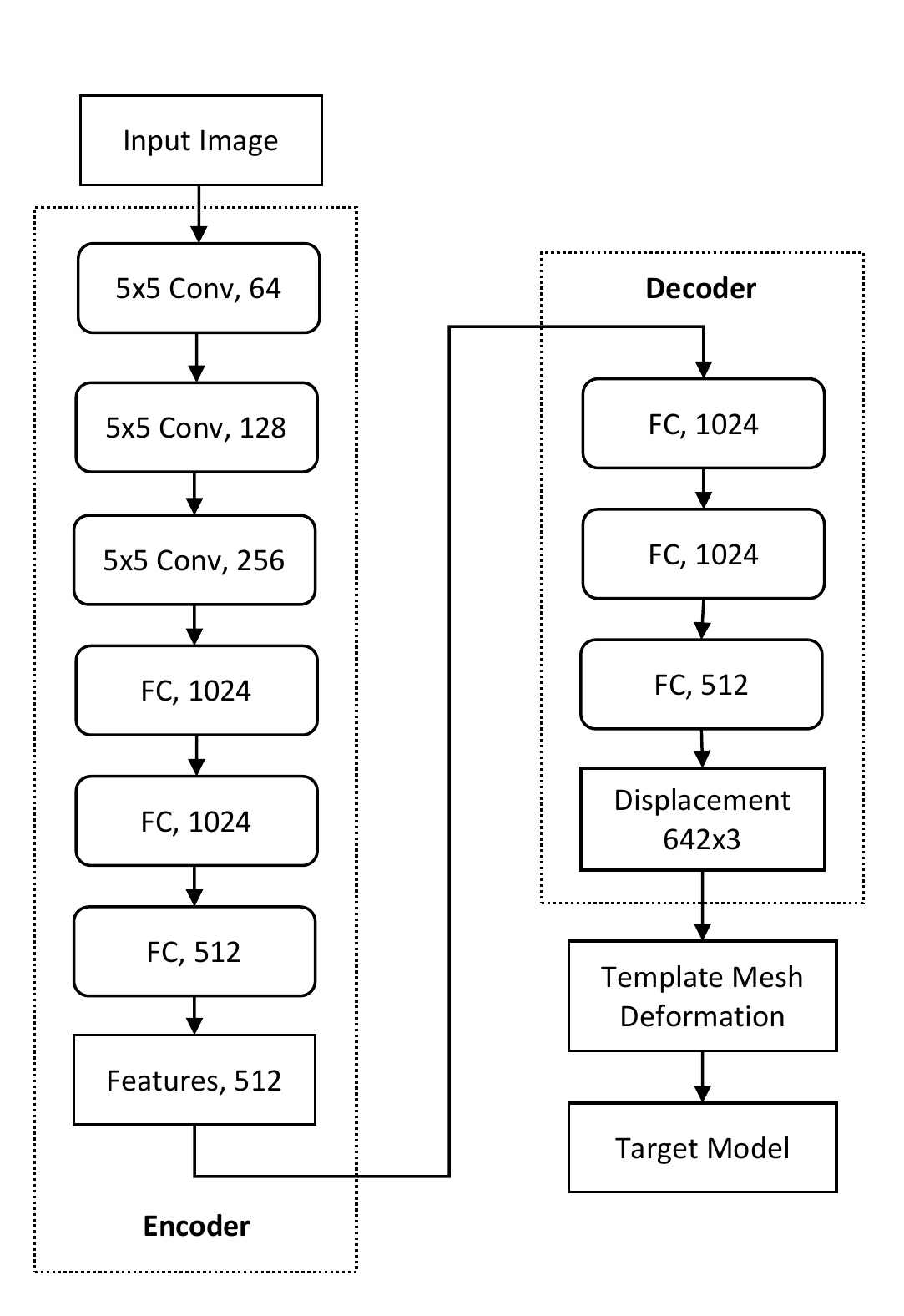}
\caption{DNN architecture.}
\label{NN_arch}
\end{figure}

Soft Rasterizer \cite{liu2019soft} is a differentiable rendering framework to train a neural network to infer 3D information from 2D images. This learning approach combined with the encoder-decoder architecture \cite{kato2018neural,yan2016perspective} can be used for mesh reconstruction of 3D models from single view image by deforming a template mesh. We employ an encoder-decoder architecture identical to \cite{liu2019soft} for single-view mesh reconstruction. The encoder is used as a feature extractor from the 2D images whereas the decoder generates the per-vertex displacement vector that deforms a template mesh into a target model based on the input 2D image. The encoder contains three convolution (Conv) and three fully connected (FC) layers and outputs a feature vector. The decoder is composed of three FC layers and outputs per-vertex displacement vector to deform a template mesh into the desired model. The detailed network structure is illustrated in Fig.~\ref{NN_arch}. We have used the dataset provided by \cite{kato2018neural}, which contains 13 categories of objects from ShapeNet \cite{chang2015shapenet}. Out of 13 categories, we have trained the DNN for two categories "Chairs" and "Tables". Each 3D model is rendered in 24 different views with image resolution of 64x64 to generate synthetic (2D images) data to train the DNN. The SR-DNN has been trained on a single NVIDIA GeForce GTX 1060 GPU and implemented using PyTorch. 

% As the training of the neural network is supervised by 2D images (silhouette) of the 3D model with multiple views instead of the 3D model, therefore the training is termed as unsupervised.

The user captures the image of a real object using HoloLens. The captured image is sent to the cloud to be accessed by the laptop. As the DNN for the mesh generation is trained using the synthetic data of 2D images rendered from the 3D models therefore to use the DNN on the real images captured through HoloLens, we need to complete the image prepossessing. To remove the background of the input image we have used an online background removal API\footnote{https://www.remove.bg/}. After removing the background, the image is cropped and resized to the image resolution of 64x64 to feed into the DNN. The 3D model generated from DNN is uploaded to the cloud to be accessed by the HoloLens and display it to the user in MR. 

\begin{figure}
\centering
\includegraphics[width=3.25in]{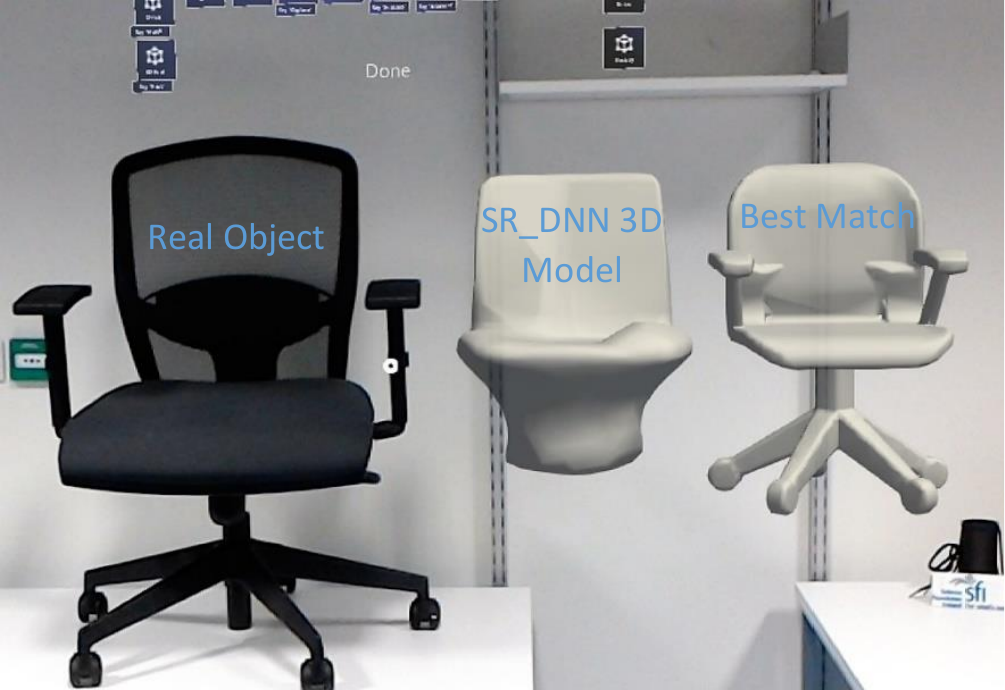}
\caption{Real object, SR-DNN generated 3D model and the best match in MR.}
\label{gen_MR}
\end{figure}

Although an active research area, the single view 3D mesh reconstruction area \cite{liu2019soft,chen2019learning,xu2019disn,jiang2020sdfdiff,mandikal20183d,zhou2020learning,wu2020pq,pinheiro2019domain,tatarchenko2019single} is still in its infancy in the context of being able to generate 3D models for AM. The Soft Rasterizer DNN (SR-DNN) reconstructs the mesh by deforming a template mesh of genus zero, therefore all 3D models generated from SR-DNN are also of the genus same as template and hence unable to match the topology of the real objects. Also, the generated 3D models are not suited for 3D printing. For this reason, we have used the 3D intersection over union (IoU) metric \cite{liu2019soft} to find the best match of the reconstructed mesh from a 3D model database which can be 3D printed. The HoloLens sends the generated 3D model to the cloud to be accessed by the laptop. The laptop then computes the 3D IoU score of the generated 3D model with all the 3D models in the database. The 3D model in the database with the maximum score (best match) is then sent to the cloud to be first accessed by the HoloLens and then displayed by the glasses to the user in MR. As an illustrative example, Fig.~\ref{gen_MR} depicts a chair, a 3D model generated using SR-DNN, and the best match shown to the user in MR using I-nteract 2.0. The model with the highest 3D IoU score is the best quantitative match but might not be the best qualitative match from the user's perspective\cite{sun2018pix3d}. Therefore we display generated 3D models with the top five scores to the user for qualitative assessment. The user can choose the best qualitative match for 3D printing.

The user can send the 3D model to the 3D printer by using the command "Print". The HoloLens sends the 3D model to the cloud to be accessed by the laptop. The laptop then converts the OBJ file of the 3D model to the G-code and then sends the G-code of the 3D model to the OctoPrint to be 3D printed. Before sending the 3D model to the 3D printer, the user can resize the 3D model to fit the physical workspace in MR. The resizing of the 3D model using I-nteract~2.0 is described in the next section.

%\IEEEpubidadjcol

\subsection{Resizing 3D models in the physical workspace}
I-nteract~2.0 provides an intuitive interface to resize a 3D model using hand motion in a MR environment. This functionality can be used to resize a 3D model according to the space in the real world. The method of resizing a 3D model is similar to the method of drawing a primitive shape described in Sec.~\ref{CSG}. The procedure implemented to resize a 3D model using I-nteract~2.0 is described below. 

\begin{enumerate}
    \item Press the "Select" virtual button using the hand without glove or use the voice command "Select" and then press on the 3D model (like pressing any virtual button) to select the 3D model. 
    \item After selecting the desired 3D model, press the "Resize" virtual button or use voice command "Resize". The index finger of the hand with glove should be open while resizing the 3D model. The width, height, and depth of the 3D model can be adjusted by moving the hand with glove in left/right (x), up/down (y), and forward/backward (z) direction respectively. Close the index finger of the hand with glove when done.
\end{enumerate}

\begin{figure}
\centering
\includegraphics[width=3.25in]{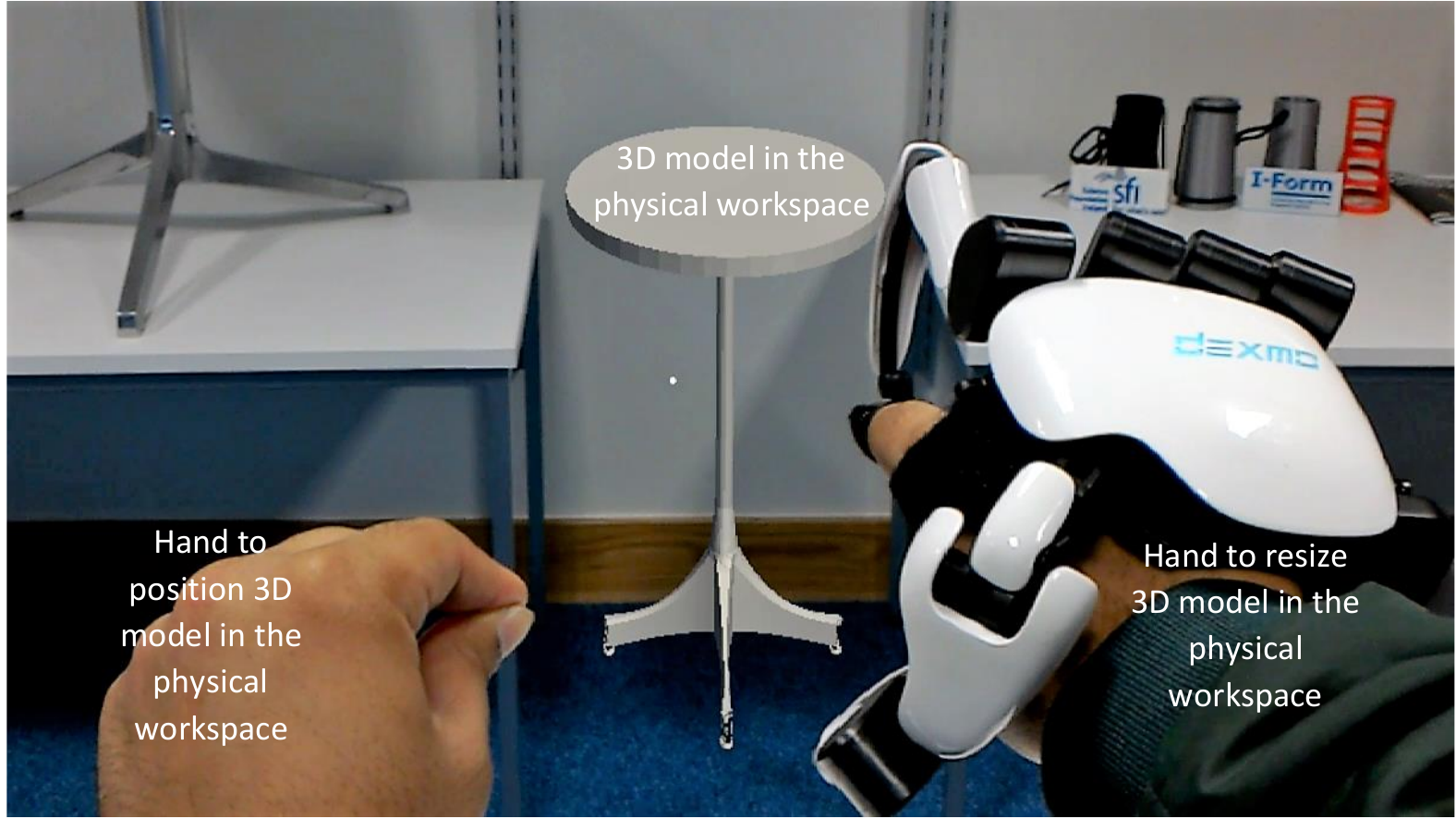}
\caption{Resizing 3D model to fit in the physical workspace using I-nteract 2.0.}
\label{fit_space}
\end{figure}

On the execution of the "Resize" command, the HoloLens records the position of the hand with glove. The HoloLens then updates (scales) the x, y and z-coordinates of the vertices of the 3D model with respect to the change in the hand position in x (left/right), y (up/down), and z (forward/backward) direction respectively. As the hand with glove will be in use while resizing the 3D model, therefore the user can use the hand without glove to position the 3D model in the physical workspace via built-in gesture (and "ManipulationHandler" script) of the HoloLens as shown in Fig.~\ref{fit_space}.

\section{Results and discussion}

\begin{figure}
\centering
\includegraphics[width=3.25in]{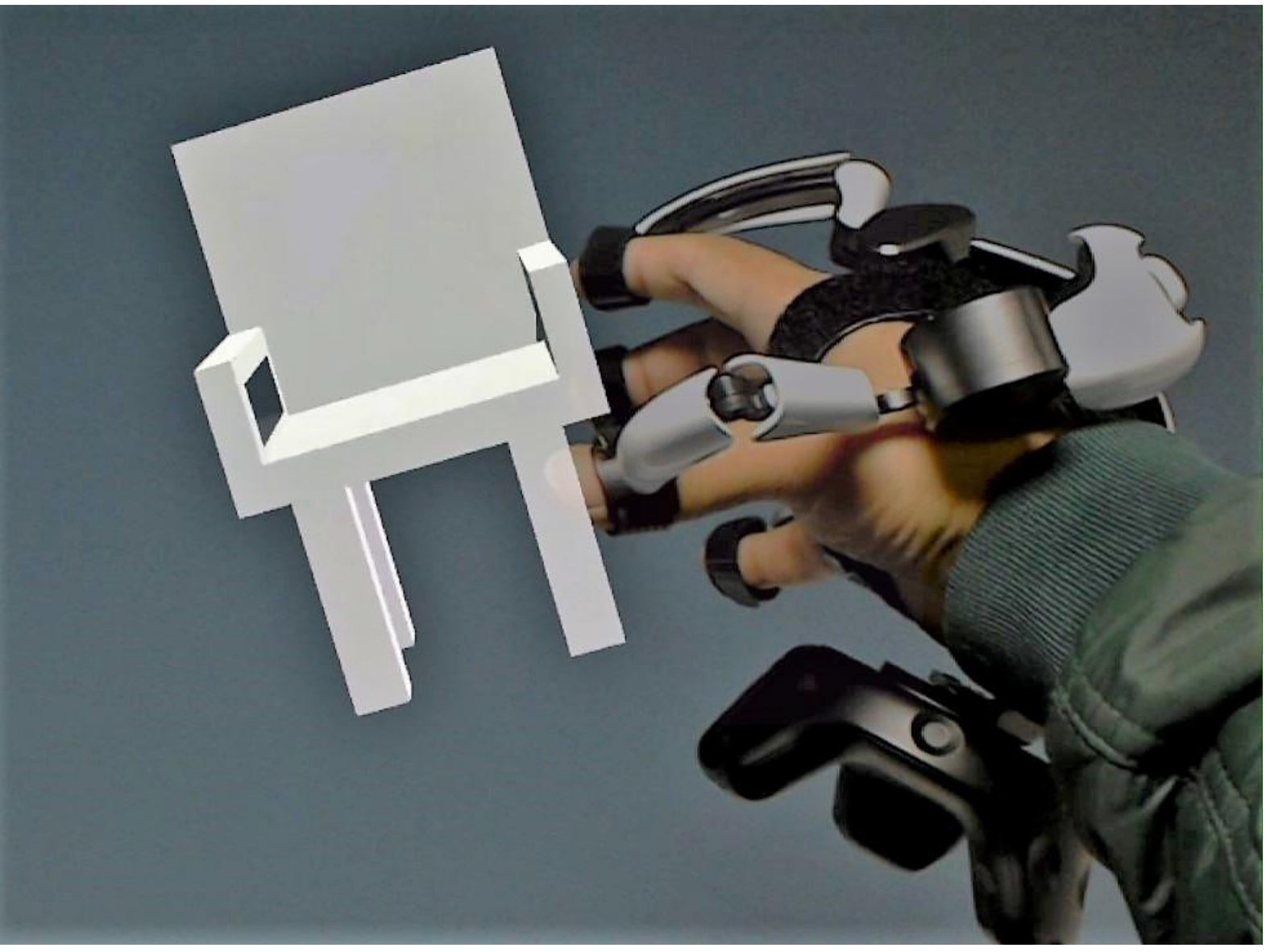}
\caption{User interacting with CSG generated chair in MR.}
\label{CSG_Hand}
\end{figure}

\begin{figure}
\centering
\includegraphics[width=2in]{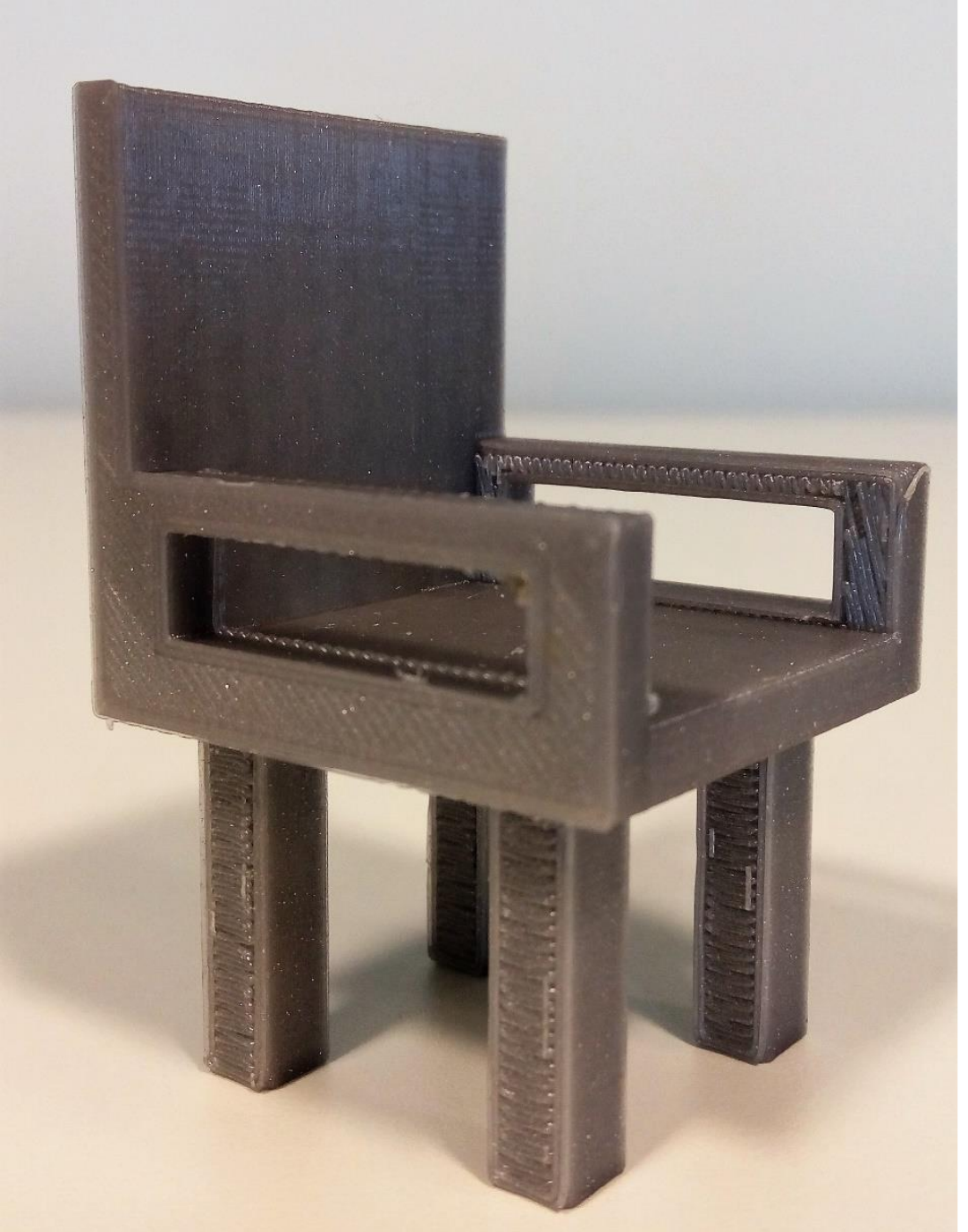}
\caption{3D print of a chair generated using CSG with I-nteract 2.0.}
\label{CSG_Print}
\end{figure}

% I-nteract 2.0 not only provides the visual feedback but also the force feedback to make the user experience with the virtual objects as physical as possible and hence making the interaction with the three dimensional virtual objects more intuitive as compared to the traditional 2D CAD interfaces. 
Figure~\ref{CSG_Hand} depicts a user interacting (translating, rotating, and getting force feedback) with the CSG generated 3D model of a chair in MR. The 3D model has been generated by applying transformations and boolean operations to primitive shapes (cube) as illustrated in Fig.~\ref{CSG_fig} in an immersive MR environment using I-nteract 2.0. The 3D print of the generated 3D model is shown in Fig.~\ref{CSG_Print}. 

% Generating a 3D model using an image is an alternative method to quickly get the model to 3D print without the need of drawing the 3D model from scratch. 

\begin{figure}
\centering
\includegraphics[width=3.25in]{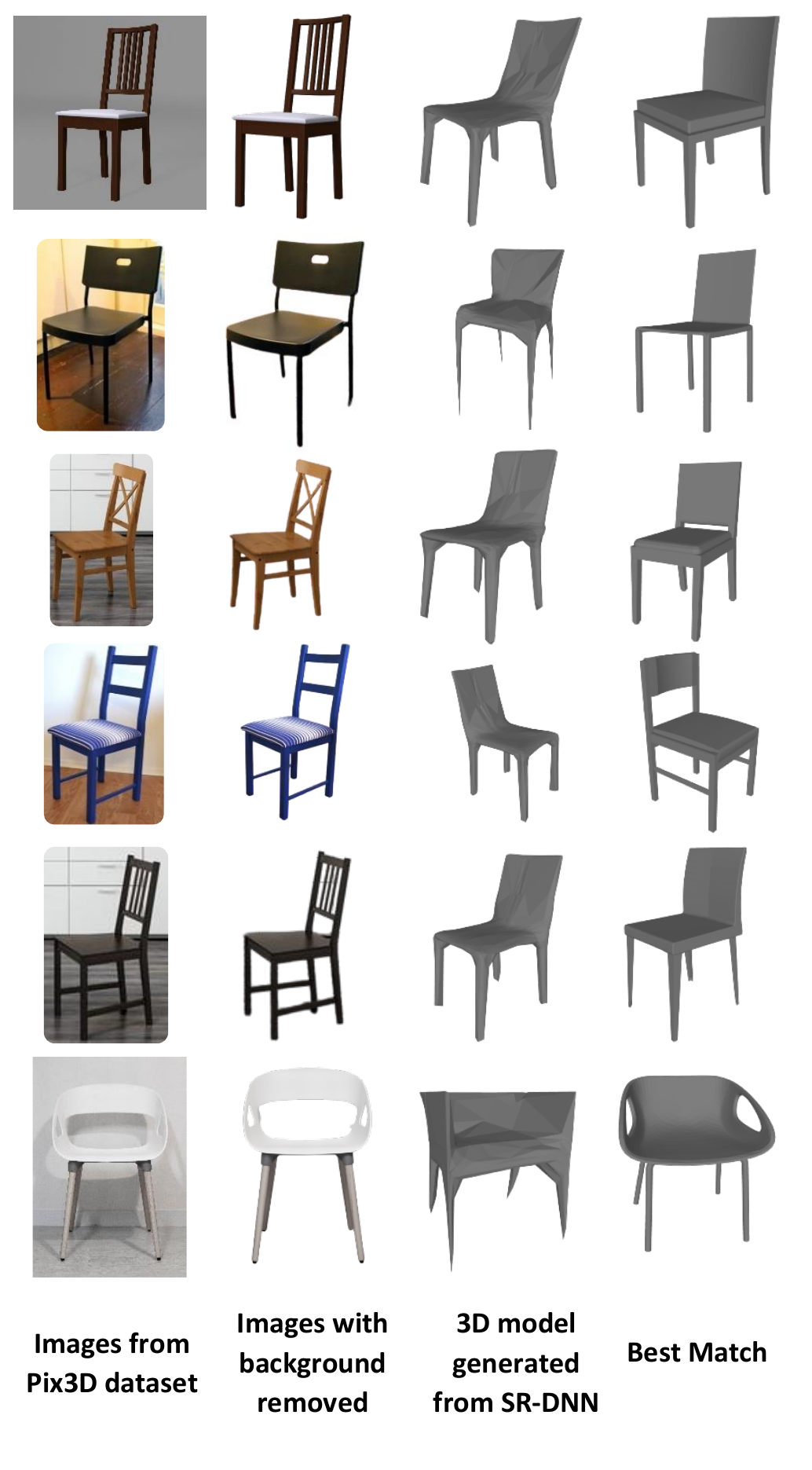}
\caption{SR-DNN tested on real images from Pix3D dataset for single-view mesh reconstruction.}
\label{Pix3D}
\end{figure}

To test the SR-DNN on real images to  generate 3D models, we have first used Pix3D \cite{sun2018pix3d} dataset. Pix3D is a dataset that consists of real images captured in diverse environments and ground-truth 3D models with nine object categories. We have tested the SR-DNN on the chair dataset and the corresponding results are shown in Fig.~\ref{Pix3D}. As the DNN is trained using the synthetic data consisting of images without any background, noise, and occlusion with multiple views rendered from the 3D model of an object therefore is not robust and requires noise free, transparent background image with a complete 2D view of the object to perform mesh reconstruction. That is why, we have tested the images from the dataset which are clear, consist of simple background, and are without any occlusion for single-view mesh reconstruction using SR-DNN. Future developments of our system will be devoted to train the DNN on challenging and realistic datasets like Pix3D \cite{sun2018pix3d}. This will improve the robustness of the DNN to extract features directly from the pictures and hence making the use of AI-based background removal API\footnote{https://www.remove.bg/} redundant.  

\begin{figure*}
\centering
\includegraphics[width=6.25in]{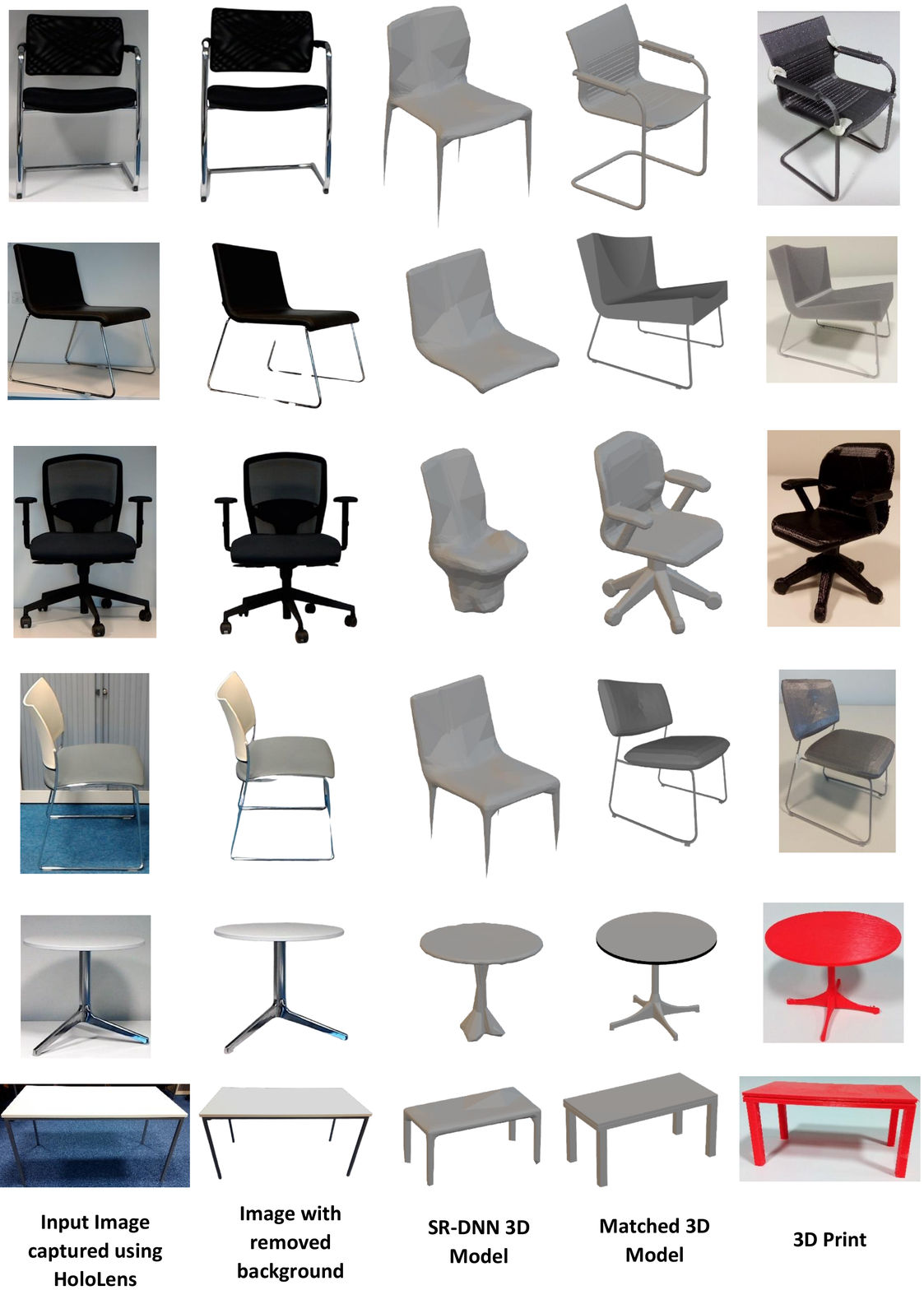}
\caption{Images, the corresponding 3D models, and 3D prints.}
\label{AI_models}
\end{figure*}

After successfully testing SR-DNN on the Pix3D dataset, we tested the SR-DNN integrated with the system (I-nteract) for 3D model generation based on a 2D image. The images captured using HoloLens, the images after removing background, the 3D models generated by the SR-DNN, the best match of the generated 3D models from the database, and their 3D prints are depicted in Fig.~\ref{AI_models}. As also mentioned above, the SR-DNN is trained using synthetic dataset which makes it non-robust with respect to images with challenging background or occlusion. This can be observed in the second row of Fig.~\ref{AI_models} because the SR-DNN was unable to generate the legs of the chair due to the distortion in the leg region of the input image induced by the removing background step. It can be seen in the third column of both Fig.~\ref{Pix3D} and Fig.~\ref{AI_models} that although SR-DNN can only generate 3D models with genus zero, it is able to faithfully infer the general 3D shapes of the objects from the 2D images.

\begin{figure}
\centering
\includegraphics[width=2in]{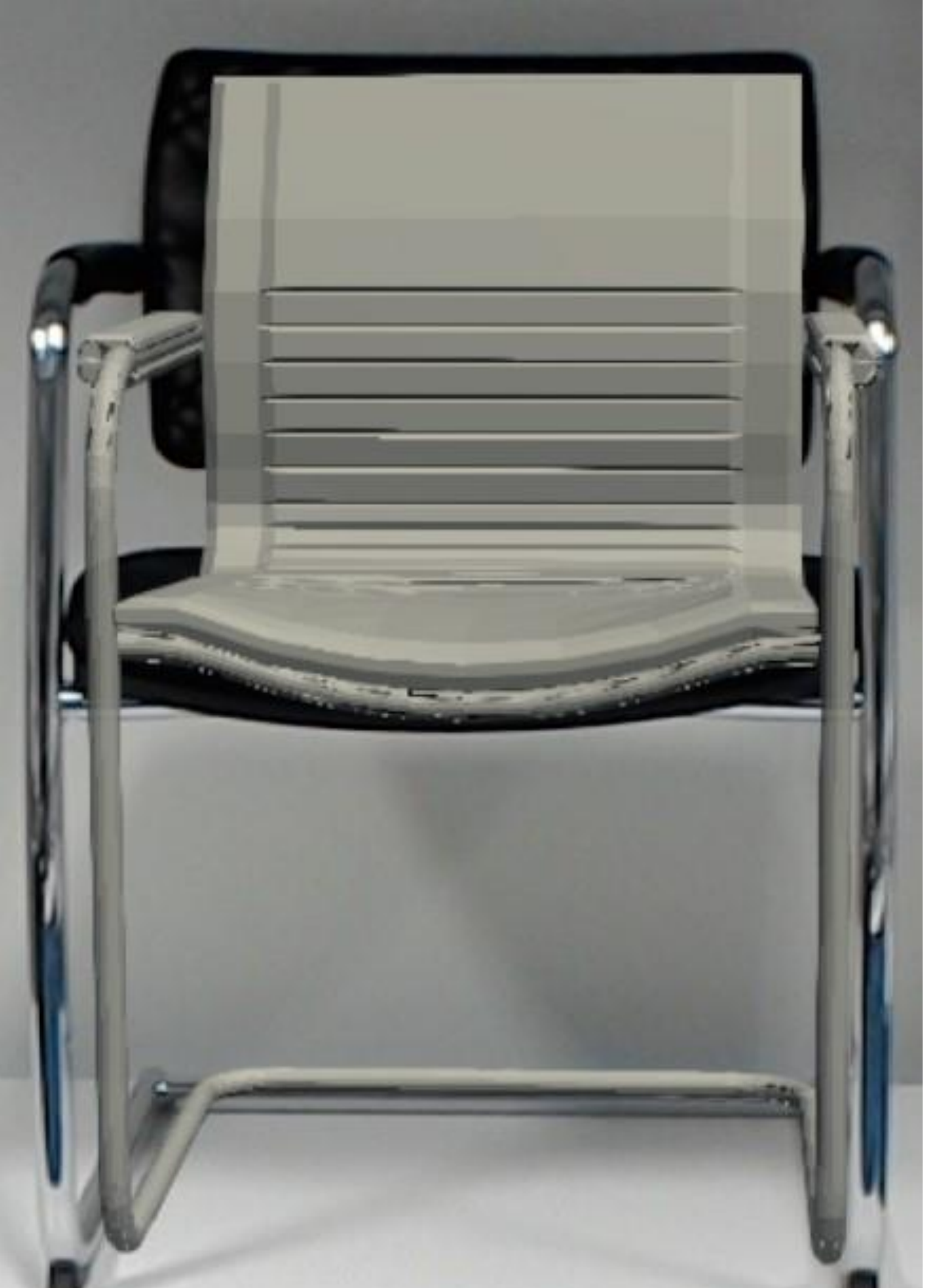}
\caption{Resizing the 3D model of a chair w.r.t a physical chair.}
\label{Resize_onChair}
\end{figure}

\begin{figure}
\centering
\includegraphics[width=2in]{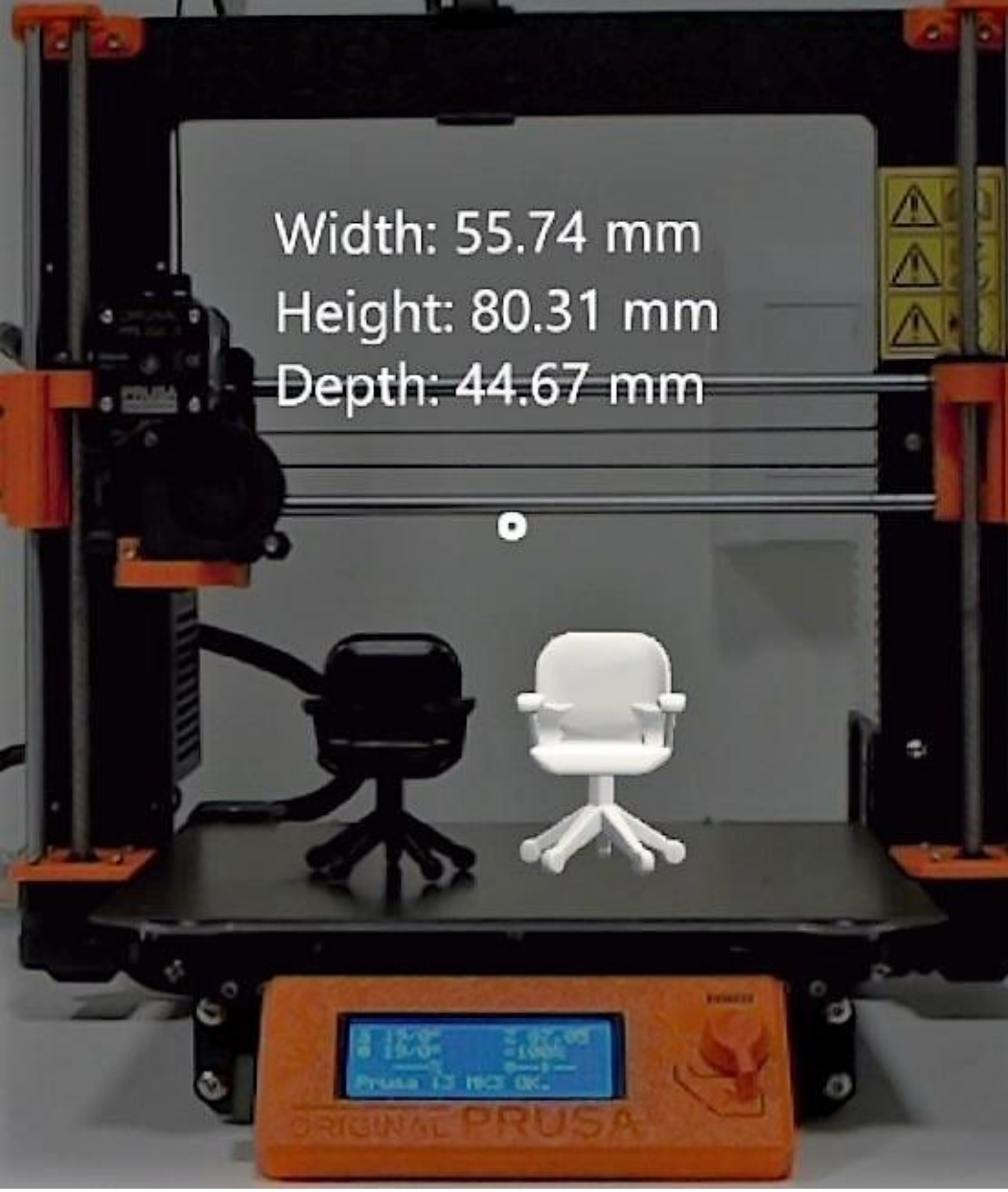}
\caption{Resized 3D model and its 3D print.}
\label{Resize_onPrinter}
\end{figure}

Fig.~\ref{Resize_onChair} shows the user resizing the matched 3D model of the chair by projecting the 3D model onto the real chair. Fig.~\ref{fit_space} depicts the user resizing the 3D model of a table to fit in a physical workspace between the two real tables. Fig.~\ref{Resize_onPrinter} shows the resized 3D model of a table onto the base plate of the 3D printer along with its 3D print. To view the dimensions of a 3D model while resizing as shown in Fig.~\ref{Resize_onPrinter} the user can execute the dimension command either via the virtual button or voice. The dimensions are computed based on the vertices positions in the OBJ file, allowing to display the maximum width, height, and depth of the 3D model. 

\begin{figure}
\centering
\includegraphics[width=2in]{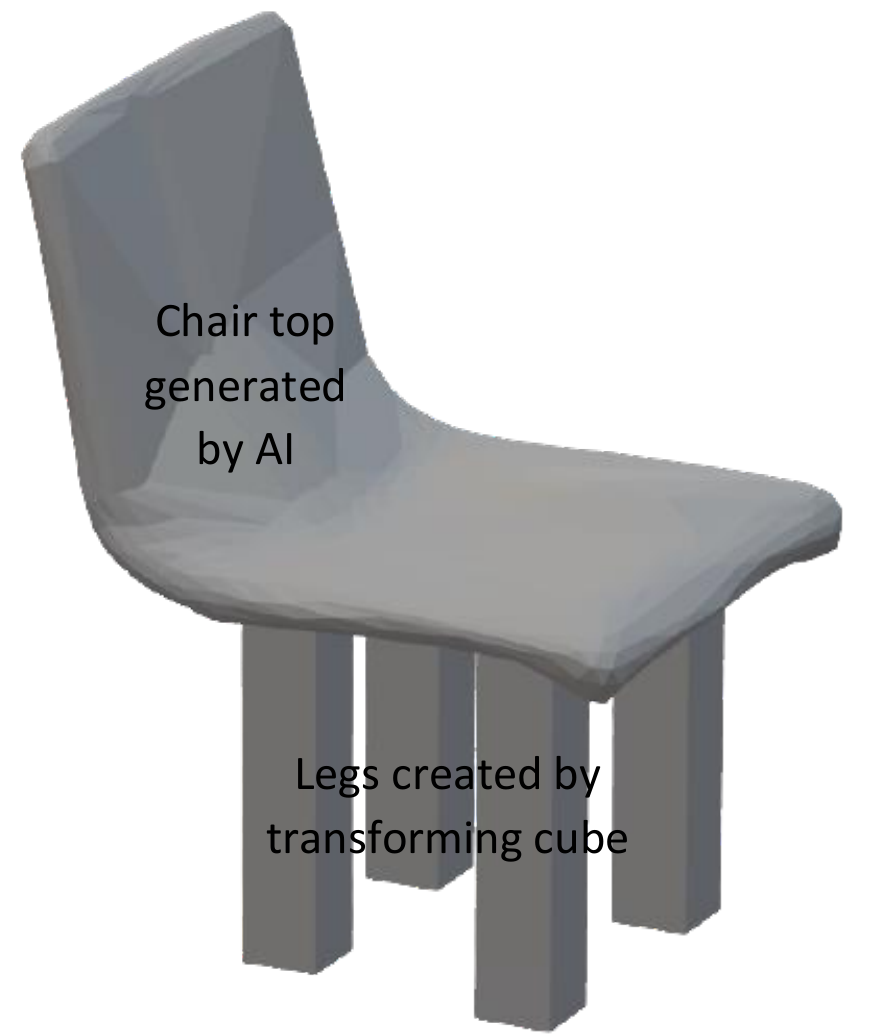}
\caption{Modified DNN generated 3D model using CSG.}
\label{CSG_AI}
\end{figure}

Future work includes enabling the user to modify the DNN generated 3D model using CSG in MR such as illustrated in Fig.~\ref{CSG_AI}. In Fig.~\ref{CSG_AI}, it can be seen that the 3D model generated from the image of a chair by SR-DNN does not have legs. Hence the 3D model is modified by adding legs using CSG. The modified model shown in Fig.~\ref{CSG_AI} has been created using Blender\footnote{https://www.blender.org/}. This kind of interface will allow the user to easily modify an existing 3D model without the need to create a 3D model from scratch. The user can capture an image of the desired object or download it from the internet to get the 3D model from the DNN and further modify it using CSG. The user can also modify an existing 3D model downloaded from the internet using CSG. The metric for finding the best match from the database (like 3D IoU) will make sure that the modified model can be 3D printed or a CAD repair API (e.g., Netfabb\footnote{https://www.autodesk.com/products/netfabb/overview}) can be integrated with the MR system to make the modified 3D model printable. Another interesting application that emerges from using generative DNN is the latent space interpolation and arithmetic~\cite{wu2016learning}. Enabling latent space interpolation in I-nteract~2.0 will allow the user to take images of two objects and generate a 3D model based on the objects in the two images. The future work will also be devoted to the use of haptic force feedback and force sensing capabilities of I-nteract to enable the user to transform the shape of a virtual object using hands in a MR environment.

\section{Conclusion}
In this paper, we presented developmental advances in the CPS named I-nteract. I-nteract provides a framework to develop intuitive and automated interfaces to allow non-technical consumers to design customized products for personal fabrication. We have introduced CAD in the VHMR system (I-nteract) by integrating it with CSG and SR-DNN to enable a novice user to design 3D models from scratch and named the developed system as I-nteract 2.0. I-nteract 2.0 also enables the user to adjust the dimensions of a virtual model with respect to the constraints in the physical workspace. The efficacy of the system has been demonstrated by generating a 3D model using CSG, then by generating 3D models of chairs and tables using a DNN from the 2D images captured via HoloLens, and finally by resizing the 3D models using hands in a MR environment with respect to the physical workspace under three scenarios. We intend to introduce additional features in our system to streamline the design process of AM using latest technological tools (such as MR, ML, robotics, IoT) in coherence to the vision of Industry 4.0. Our objective is to develop an intuitive and user-friendly interface that not only enables personal fabrication by allowing non-technical user to generate and modify 3D models in shape and size as well as by automating the parts of the design process that require expert knowledge but also provides effective means of monitoring the AM process to improve the build quality of the product.

% or


% \appendices

% % you can choose not to have a title for an appendix
% % if you want by leaving the argument blank
% \section{}

%\section*{Acknowledgment}


\ifCLASSOPTIONcaptionsoff
  \newpage
\fi



\bibliographystyle{IEEEtran}
\bibliography{IEEEabrv,Library_human_machine_interaction}

\begin{thebibliography}{10}
\providecommand{\url}[1]{#1}
\csname url@samestyle\endcsname
\providecommand{\newblock}{\relax}
\providecommand{\bibinfo}[2]{#2}
\providecommand{\BIBentrySTDinterwordspacing}{\spaceskip=0pt\relax}
\providecommand{\BIBentryALTinterwordstretchfactor}{4}
\providecommand{\BIBentryALTinterwordspacing}{\spaceskip=\fontdimen2\font plus
\BIBentryALTinterwordstretchfactor\fontdimen3\font minus
  \fontdimen4\font\relax}
\providecommand{\BIBforeignlanguage}[2]{{%
\expandafter\ifx\csname l@#1\endcsname\relax
\typeout{** WARNING: IEEEtran.bst: No hyphenation pattern has been}%
\typeout{** loaded for the language `#1'. Using the pattern for}%
\typeout{** the default language instead.}%
\else
\language=\csname l@#1\endcsname
\fi
#2}}
\providecommand{\BIBdecl}{\relax}
\BIBdecl

\bibitem{shorten2020industry}
R.~Shorten, J.~Oliver, D.~Clayton, A.~Malik, and H.~Lhachemi, ``Industry 4.0
  and the sharing economy,'' in \emph{Analytics for the Sharing Economy:
  Mathematics, Engineering and Business Perspectives}.\hskip 1em plus 0.5em
  minus 0.4em\relax Springer, 2020, pp. 335--345.

\bibitem{RN118}
M.~Cotteleer and J.~Joyce, ``{3D} opportunity: Additive manufacturing paths to
  performance, innovation, and growth,'' \emph{Deloitte Review}, 2014.

\bibitem{RN6}
T.~Stock and G.~Seliger, ``Opportunities of sustainable manufacturing in
  industry 4.0,'' \emph{Procedia CIRP}, vol.~40, pp. 536--541, 2016.

\bibitem{attaran2017rise}
M.~Attaran, ``The rise of 3-{D} printing: The advantages of additive
  manufacturing over traditional manufacturing,'' \emph{Business Horizons},
  vol.~60, no.~5, pp. 677--688, 2017.

\bibitem{RN87}
B.~Fröhlich and J.~Plate, ``The cubic mouse: a new device for
  three-dimensional input,'' in \emph{Proceedings of the SIGCHI conference on
  Human Factors in Computing Systems}.\hskip 1em plus 0.5em minus 0.4em\relax
  ACM, 2000, Conference Proceedings, pp. 526--531.

\bibitem{RN53}
S.~Mueller, ``{3D} printing for human-computer interaction,''
  \emph{Interactions}, vol.~24, no.~5, pp. 76--79, 2017.

\bibitem{lhachemi2019augmented}
H.~Lhachemi, A.~Malik, and R.~Shorten, ``Augmented reality, cyber-physical
  systems, and feedback control for additive manufacturing: a review,''
  \emph{IEEE Access}, vol.~7, pp. 50\,119--50\,135, 2019.

\bibitem{malik2020nteract}
A.~Malik, H.~Lhachemi, and R.~Shorten, ``I-nteract: A cyber-physical system for
  real-time interaction with physical and virtual objects using mixed reality
  technologies for additive manufacturing,'' \emph{IEEE Access}, 2020.

\bibitem{RN1}
C.~Weichel, M.~Lau, D.~Kim, N.~Villar, and H.~W. Gellersen, ``{MixFab}: a
  mixed-reality environment for personal fabrication,'' in \emph{Proceedings of
  the SIGCHI Conference on Human Factors in Computing Systems}.\hskip 1em plus
  0.5em minus 0.4em\relax ACM, 2014, Conference Proceedings, pp. 3855--3864.

\bibitem{RN13}
O.~Hilliges, D.~Kim, S.~Izadi, M.~Weiss, and A.~Wilson, ``{HoloDesk}: direct 3d
  interactions with a situated see-through display,'' in \emph{Proceedings of
  the SIGCHI Conference on Human Factors in Computing Systems}.\hskip 1em plus
  0.5em minus 0.4em\relax ACM, 2012, Conference Proceedings, pp. 2421--2430.

\bibitem{RN45}
C.-H. Hsu, W.-H. Cheng, and K.-L. Hua, ``{HoloTabletop}: an anamorphic illusion
  interactive holographic-like tabletop system,'' \emph{Multimedia Tools and
  Applications}, vol.~76, no.~7, pp. 9245--9264, 2017.

\bibitem{malik2019}
A.~Malik, H.~Lhachemi, J.~Ploennigs, A.~Ba, and R.~Shorten, ``An application of
  3{D} model reconstruction and augmented reality for real-time monitoring of
  additive manufacturing,'' \emph{Procedia CIRP}, vol.~81, pp. 346--351, 2019.

\bibitem{RN77}
A.~Ceruti, A.~Liverani, and T.~Bombardi, ``Augmented vision and interactive
  monitoring in {3D} printing process,'' \emph{International Journal on
  Interactive Design and Manufacturing (IJIDeM)}, vol.~11, no.~2, pp. 385--395,
  2017.

\bibitem{klahn2015design}
C.~Klahn, B.~Leutenecker, and M.~Meboldt, ``Design strategies for the process
  of additive manufacturing,'' \emph{Procedia Cirp}, vol.~36, pp. 230--235,
  2015.

\bibitem{requicha1977constructive}
A.~A. Requicha and H.~B. Voelcker, ``Constructive solid geometry,'' 1977.

\bibitem{liu2019soft}
S.~Liu, T.~Li, W.~Chen, and H.~Li, ``Soft rasterizer: A differentiable renderer
  for image-based 3d reasoning,'' in \emph{Proceedings of the IEEE
  International Conference on Computer Vision}, 2019, pp. 7708--7717.

\bibitem{kato2018neural}
H.~Kato, Y.~Ushiku, and T.~Harada, ``Neural 3d mesh renderer,'' in
  \emph{Proceedings of the IEEE Conference on Computer Vision and Pattern
  Recognition}, 2018, pp. 3907--3916.

\bibitem{yan2016perspective}
X.~Yan, J.~Yang, E.~Yumer, Y.~Guo, and H.~Lee, ``Perspective transformer nets:
  Learning single-view 3d object reconstruction without 3d supervision,'' in
  \emph{Advances in neural information processing systems}, 2016, pp.
  1696--1704.

\bibitem{sinha2017surfnet}
A.~Sinha, A.~Unmesh, Q.~Huang, and K.~Ramani, ``Surfnet: Generating 3d shape
  surfaces using deep residual networks,'' in \emph{Proceedings of the IEEE
  conference on computer vision and pattern recognition}, 2017, pp. 6040--6049.

\bibitem{lin2017learning}
C.-H. Lin, C.~Kong, and S.~Lucey, ``Learning efficient point cloud generation
  for dense 3d object reconstruction,'' \emph{arXiv preprint arXiv:1706.07036},
  2017.

\bibitem{dosovitskiy2016learning}
A.~Dosovitskiy, J.~T. Springenberg, M.~Tatarchenko, and T.~Brox, ``Learning to
  generate chairs, tables and cars with convolutional networks,'' \emph{IEEE
  transactions on pattern analysis and machine intelligence}, vol.~39, no.~4,
  pp. 692--705, 2016.

\bibitem{wang2017shape}
W.~Wang, Q.~Huang, S.~You, C.~Yang, and U.~Neumann, ``Shape inpainting using 3d
  generative adversarial network and recurrent convolutional networks,'' in
  \emph{Proceedings of the IEEE International Conference on Computer Vision},
  2017, pp. 2298--2306.

\bibitem{wu2016learning}
J.~Wu, C.~Zhang, T.~Xue, B.~Freeman, and J.~Tenenbaum, ``Learning a
  probabilistic latent space of object shapes via 3d generative-adversarial
  modeling,'' in \emph{Advances in neural information processing systems},
  2016, pp. 82--90.

\bibitem{fan2017point}
H.~Fan, H.~Su, and L.~J. Guibas, ``A point set generation network for 3d object
  reconstruction from a single image,'' in \emph{Proceedings of the IEEE
  conference on computer vision and pattern recognition}, 2017, pp. 605--613.

\bibitem{gadelha20173d}
M.~Gadelha, S.~Maji, and R.~Wang, ``3d shape induction from 2d views of
  multiple objects,'' in \emph{2017 International Conference on 3D Vision
  (3DV)}.\hskip 1em plus 0.5em minus 0.4em\relax IEEE, 2017, pp. 402--411.

\bibitem{RN91}
K.~Huo and K.~Ramani, ``Window-shaping: {3D} design ideation by creating on,
  borrowing from, and looking at the physical world,'' in \emph{Proceedings of
  the Tenth International Conference on Tangible, Embedded, and Embodied
  Interaction}.\hskip 1em plus 0.5em minus 0.4em\relax ACM, 2017, Conference
  Proceedings, pp. 37--45.

\bibitem{RN9}
H.~Benko, R.~Jota, and A.~Wilson, ``{MirageTable}: freehand interaction on a
  projected augmented reality tabletop,'' in \emph{Proceedings of the SIGCHI
  conference on human factors in computing systems}.\hskip 1em plus 0.5em minus
  0.4em\relax ACM, 2012, Conference Proceedings, pp. 199--208.

\bibitem{RN22}
S.~Schkolne, M.~Pruett, and P.~Schröder, ``Surface drawing: creating organic
  {3D} shapes with the hand and tangible tools,'' in \emph{Proceedings of the
  SIGCHI conference on Human factors in computing systems}.\hskip 1em plus
  0.5em minus 0.4em\relax ACM, 2001, Conference Proceedings, pp. 261--268.

\bibitem{RN18}
I.~Llamas, B.~Kim, J.~Gargus, J.~Rossignac, and C.~D. Shaw, ``Twister: a
  space-warp operator for the two-handed editing of {3D} shapes,'' \emph{ACM
  transactions on graphics (TOG)}, vol.~22, no.~3, pp. 663--668, 2003.

\bibitem{RN16}
D.~Kim, O.~Hilliges, S.~Izadi, A.~D. Butler, J.~Chen, I.~Oikonomidis, and
  P.~Olivier, ``Digits: freehand {3D} interactions anywhere using a wrist-worn
  gloveless sensor,'' in \emph{Proceedings of the 25th annual ACM symposium on
  User interface software and technology}.\hskip 1em plus 0.5em minus
  0.4em\relax ACM, 2012, Conference Proceedings, pp. 167--176.

\bibitem{RN72}
H.~Benko, C.~Holz, M.~Sinclair, and E.~Ofek, ``Normaltouch and texturetouch:
  High-fidelity 3d haptic shape rendering on handheld virtual reality
  controllers,'' in \emph{Proceedings of the 29th Annual Symposium on User
  Interface Software and Technology}.\hskip 1em plus 0.5em minus 0.4em\relax
  ACM, 2016, Conference Proceedings, pp. 717--728.

\bibitem{meng2020machine}
L.~Meng, B.~McWilliams, W.~Jarosinski, H.-Y. Park, Y.-G. Jung, J.~Lee, and
  J.~Zhang, ``Machine learning in additive manufacturing: A review,''
  \emph{JOM}, pp. 1--15, 2020.

\bibitem{oh2019deep}
S.~Oh, Y.~Jung, S.~Kim, I.~Lee, and N.~Kang, ``Deep generative design:
  Integration of topology optimization and generative models,'' \emph{Journal
  of Mechanical Design}, vol. 141, no.~11, 2019.

\bibitem{oh2018design}
S.~Oh, Y.~Jung, I.~Lee, and N.~Kang, ``Design automation by integrating
  generative adversarial networks and topology optimization,'' in
  \emph{International Design Engineering Technical Conferences and Computers
  and Information in Engineering Conference}, vol. 51753.\hskip 1em plus 0.5em
  minus 0.4em\relax American Society of Mechanical Engineers, 2018, p.
  V02AT03A008.

\bibitem{kingma2013auto}
D.~P. Kingma and M.~Welling, ``Auto-encoding variational bayes,'' \emph{arXiv
  preprint arXiv:1312.6114}, 2013.

\bibitem{goodfellow2014generative}
I.~Goodfellow, J.~Pouget-Abadie, M.~Mirza, B.~Xu, D.~Warde-Farley, S.~Ozair,
  A.~Courville, and Y.~Bengio, ``Generative adversarial nets,'' in
  \emph{Advances in neural information processing systems}, 2014, pp.
  2672--2680.

\bibitem{wang2018pixel2mesh}
N.~Wang, Y.~Zhang, Z.~Li, Y.~Fu, W.~Liu, and Y.-G. Jiang, ``Pixel2mesh:
  Generating 3d mesh models from single rgb images,'' in \emph{Proceedings of
  the European Conference on Computer Vision (ECCV)}, 2018, pp. 52--67.

\bibitem{CSG_Unity}
W.~Evan, ``A c\# port of csg.js for use in the unity game engine.''
  \url{https://github.com/karl-/pb\_CSG}, 2020.

\bibitem{chang2015shapenet}
A.~X. Chang, T.~Funkhouser, L.~Guibas, P.~Hanrahan, Q.~Huang, Z.~Li,
  S.~Savarese, M.~Savva, S.~Song, H.~Su \emph{et~al.}, ``Shapenet: An
  information-rich 3d model repository,'' \emph{arXiv preprint
  arXiv:1512.03012}, 2015.

\bibitem{chen2019learning}
Z.~Chen and H.~Zhang, ``Learning implicit fields for generative shape
  modeling,'' in \emph{Proceedings of the IEEE Conference on Computer Vision
  and Pattern Recognition}, 2019, pp. 5939--5948.

\bibitem{xu2019disn}
Q.~Xu, W.~Wang, D.~Ceylan, R.~Mech, and U.~Neumann, ``Disn: Deep implicit
  surface network for high-quality single-view 3d reconstruction,'' in
  \emph{Advances in Neural Information Processing Systems}, 2019, pp. 492--502.

\bibitem{jiang2020sdfdiff}
Y.~Jiang, D.~Ji, Z.~Han, and M.~Zwicker, ``Sdfdiff: Differentiable rendering of
  signed distance fields for 3d shape optimization,'' in \emph{Proceedings of
  the IEEE/CVF Conference on Computer Vision and Pattern Recognition}, 2020,
  pp. 1251--1261.

\bibitem{mandikal20183d}
P.~Mandikal, K.~Navaneet, M.~Agarwal, and R.~V. Babu, ``{3D-LMNet}: Latent
  embedding matching for accurate and diverse {3D} point cloud reconstruction
  from a single image,'' \emph{arXiv preprint arXiv:1807.07796}, 2018.

\bibitem{zhou2020learning}
Y.~Zhou, S.~Liu, and Y.~Ma, ``Learning to detect {3D} reflection symmetry for
  single-view reconstruction,'' \emph{arXiv preprint arXiv:2006.10042}, 2020.

\bibitem{wu2020pq}
R.~Wu, Y.~Zhuang, K.~Xu, H.~Zhang, and B.~Chen, ``Pq-net: A generative part
  seq2seq network for {3D} shapes,'' in \emph{Proceedings of the IEEE/CVF
  Conference on Computer Vision and Pattern Recognition}, 2020, pp. 829--838.

\bibitem{pinheiro2019domain}
P.~O. Pinheiro, N.~Rostamzadeh, and S.~Ahn, ``Domain-adaptive single-view 3d
  reconstruction,'' in \emph{Proceedings of the IEEE International Conference
  on Computer Vision}, 2019, pp. 7638--7647.

\bibitem{tatarchenko2019single}
M.~Tatarchenko, S.~R. Richter, R.~Ranftl, Z.~Li, V.~Koltun, and T.~Brox, ``What
  do single-view 3d reconstruction networks learn?'' in \emph{Proceedings of
  the IEEE Conference on Computer Vision and Pattern Recognition}, 2019, pp.
  3405--3414.

\bibitem{sun2018pix3d}
X.~Sun, J.~Wu, X.~Zhang, Z.~Zhang, C.~Zhang, T.~Xue, J.~B. Tenenbaum, and W.~T.
  Freeman, ``Pix3d: Dataset and methods for single-image 3d shape modeling,''
  in \emph{Proceedings of the IEEE Conference on Computer Vision and Pattern
  Recognition}, 2018, pp. 2974--2983.

\end{thebibliography}


\begin{IEEEbiography}[{\includegraphics[width=1in,height=1.25in,clip,keepaspectratio]{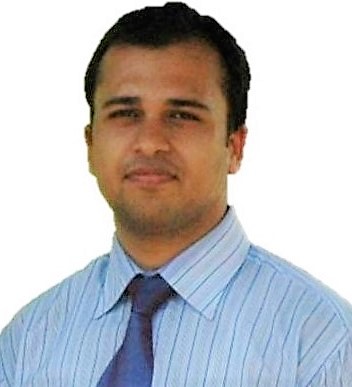}}]{Ammar Malik} received his Bachelor’s degree in Electrical Engineering and Master’s degree in Systems Engineering from Pakistan Institute of Engineering \& Applied Sciences (PIEAS), Islamabad, Pakistan. He worked as Control Systems Design Lab Engineer at Engineering Universities and as Electrical Engineer for Electromechanical Service providers in Abu Dhabi, UAE. He is currently pursuing the Ph.D. degree with the School of Electrical and Electronic Engineering at University College Dublin, Ireland. His research interests include cyber-physical systems, human-machine interaction, artificial intelligence, robotics, computer vision, data-driven control and learning systems.
\end{IEEEbiography}

\begin{IEEEbiography}[{\includegraphics[width=1in,height=1.25in,clip,keepaspectratio]{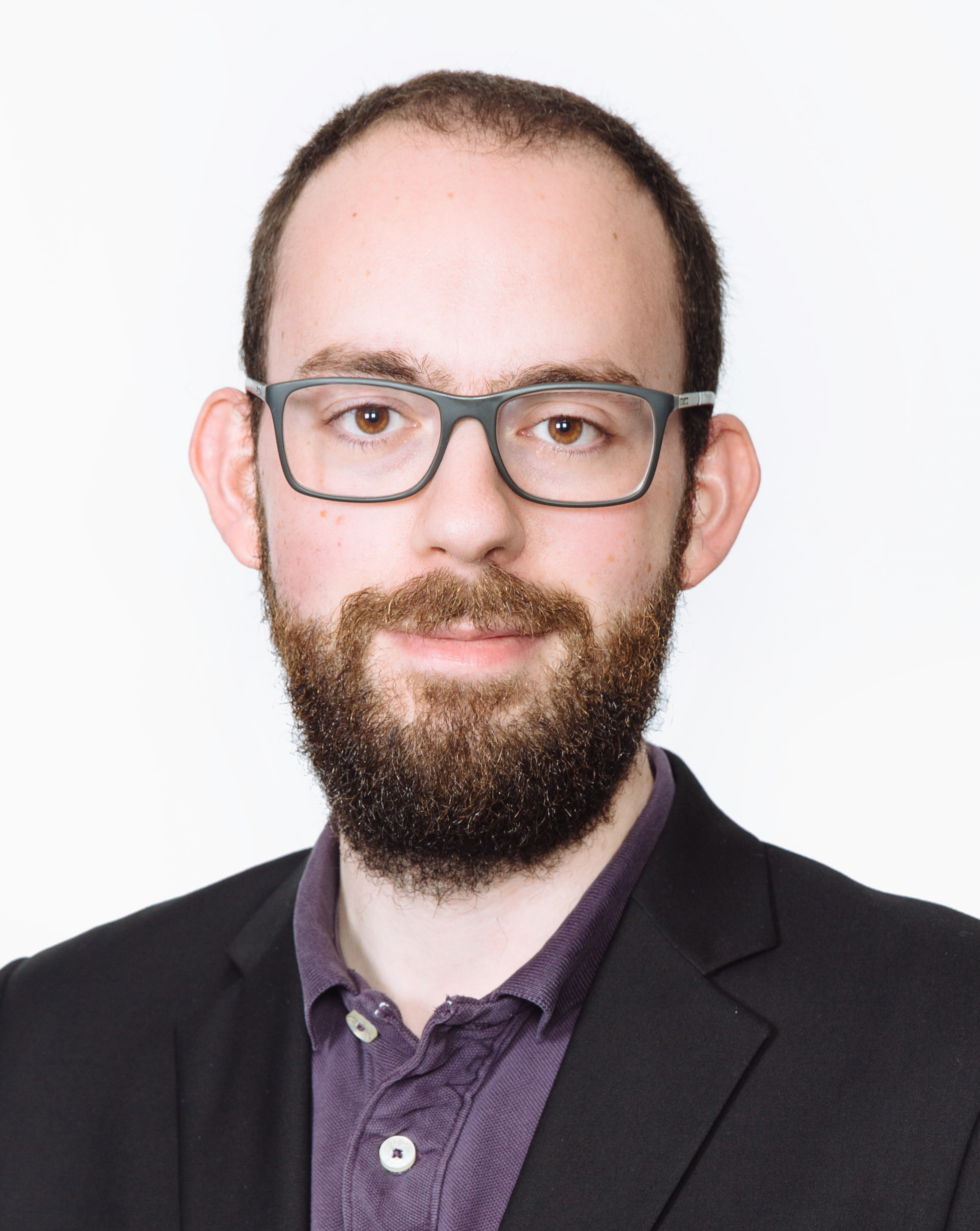}}]{Hugo Lhachemi} received a four-year university degree in mathematics from Université Claude Bernard Lyon I, France, in 2011, an engineering degree from Ecole Centrale de Lyon, France, in 2013, and a M.Sc. degree in Aerospace Engineering from Polytechnique Montreal, Canada, in 2013. He received his Ph.D. degree in Electrical Engineering from Polytechnique Montreal in 2017. He was postdoctoral fellow in Automation and Control at University College Dublin, Ireland, from 2018 to 2020. Since 2020, he is an Associate Professor of System Control at CentraleSupélec, France. His main research interests include nonlinear control, infinite dimensional systems, and their applications to aerospace and cyber-physical systems.

\end{IEEEbiography}

%\newpage

\begin{IEEEbiography}[{\includegraphics[width=1in,height=1.25in,clip,keepaspectratio]{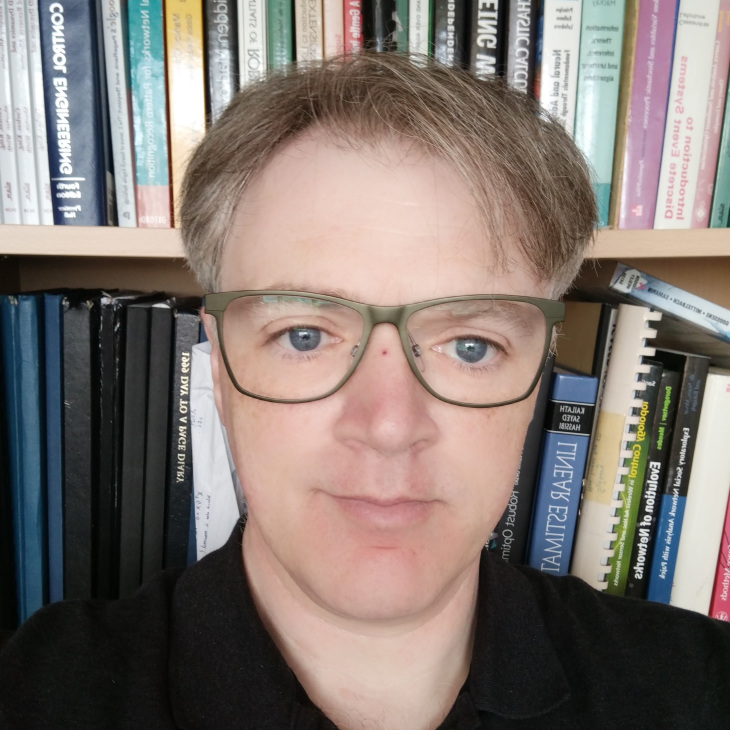}}]{Robert Shorten} was a Co-Founder of the Hamilton Institute, Maynooth University. He led the Optimisation and Control Team, IBM Research Smart Cities Research Lab, Dublin, Ireland. He has been a Visiting Professor with TU Berlin and a Research Visitor with Yale University and Technion. He is currently professor of Cyber-physical Systems at the Dyson School of Engineering Design at Imperial College London. He is also with University College Dublin. He is a co-author of the recently published books AIMD Dynamics and Distributed Resource Allocation (SIAM, 2016) and Electric and Plug-in Vehicle Networks: Optimisation and Control (CRC Press, Taylor and Francis Group, 2017). He is the Irish Member of the European Union Control Association assembly and a member of the IEEE Control Systems Society Technical Group on Smart Cities, the IFAC Technical Committee for Automotive Control, and the IFAC Technical Committee for Discrete Event and Hybrid Systems.

\end{IEEEbiography}


%\vfill


\end{document}